\documentclass[twocolumn,twocolappendix]{aastex631}

\usepackage{amsmath}
\usepackage{lipsum}
\usepackage{bm}
\usepackage{verbatim}
\usepackage{xcolor}
\usepackage{soul}
\usepackage{chngcntr}
\usepackage{ifdraft}

\usepackage{natbib}
\ifdraft{\renewcommand{\includegraphics}{\relax}}{\relax}

\definecolor{royalblue}{RGB}{65,105,225}
\definecolor{royalred}{RGB}{255,40,0}

\newcommand{\sups}{\sigma_{\rm up}}
\newcommand{\rins}{r_{\rm in}}
\newcommand{\rups}{r_{\rm up}}
\newcommand{\Rlg}{\mathcal{R}}
\newcommand{\Blg}{\mathcal{B}}

\newcommand\reone[1]{{#1}}
\newcommand\retwo[1]{{#1}}

\shorttitle{Plasmoid compression}
\shortauthors{Hakobyan et al.}

\graphicspath{{./}{figures/}}

\begin{document}

\title{Secondary Energization in Compressing Plasmoids during Magnetic Reconnection}
\correspondingauthor{Hayk Hakobyan}
\email{hakobyan@astro.princeton.edu}

\author{Hayk Hakobyan}
\affiliation{Department of Astrophysical Sciences, Princeton University, Princeton, NJ 08544, USA}

\author{Maria Petropoulou}
\affiliation{Department of Astrophysical Sciences, Princeton University, Princeton, NJ 08544, USA}

\author{Anatoly Spitkovsky}
\affiliation{Department of Astrophysical Sciences, Princeton University, Princeton, NJ 08544, USA}

\author{Lorenzo Sironi}
\affiliation{Department of Astronomy, Columbia University, New York, NY 10027, USA}

\begin{abstract}

Plasmoids -- magnetized quasi-circular structures formed self-consistently in reconnecting current sheets -- were previously considered to be the graveyards of energetic particles. In this paper, we demonstrate the important role of plasmoids in shaping the particle energy spectrum in relativistic reconnection (i.e., with upstream magnetization $\sups\gg 1$). Using 2D particle-in-cell simulations in pair plasmas with $\sups=10$ and $100$, we study a secondary particle energization process that takes place inside compressing plasmoids. We demonstrate that plasmoids grow in time, while their interiors compress, amplifying the internal magnetic field. The magnetic field felt by particles injected in an isolated plasmoid increases linearly with time, which leads to particle energization as a result of magnetic moment conservation. For particles injected with a power-law distribution function, this energization process acts in such a way that the shape of the injected power law is conserved, while producing an additional nonthermal tail $f(E)\propto E^{-3}$ at higher energies, followed by an exponential cutoff. The cutoff energy, which increases with time as $E_{\rm cut}\propto\sqrt{t}$, can greatly exceed $\sups m_e c^2$. We analytically predict the secondary acceleration timescale and the shape of the emerging particle energy spectrum, which can be of major importance in certain astrophysical systems, such as blazar jets.

\end{abstract}

\keywords{magnetic reconnection -- radiation mechanisms: non-thermal -- pulsars: general -- galaxies: jets}


%
%
%
%

\section{Introduction} \label{sec:intro}

Magnetic reconnection is a very efficient and rapid mechanism of tapping magnetic field energy in astrophysical environments. In recent decades this phenomenon has been studied extensively with numerical techniques varying from resistive magnetohydrodynamics (MHD) \citep{2005PhRvL..95w5003L, 2010PhPl...17f2104H} to kinetic particle-in-cell (PIC) algorithms \citep[e.g.,][]{2001ApJ...562L..63Z, 2012ApJ...750..129B, 2014PhRvL.113o5005G, 2014ApJ...783L..21S}. Systems of two plane-parallel magnetic field regions with opposite polarities separated by a current layer are thought to serve as good localized analogs of  larger-scale astrophysical systems. PIC simulations of such regions in the magnetically dominated relativistic regime, when the available magnetic field energy greatly exceeds the plasma energy, have been studied in the past decade. These simulations (in both in two and three dimensions) have shown that relativistic magnetic reconnection produces extended nonthermal particle energy spectra, which can usually be described by a power law with a high-energy exponential cutoff, namely, ${\mathrm d}N/{\mathrm d}E \propto E^{p} e^{-E/E_{\rm cut}}$. The power-law index $p$ is found to depend on the plasma magnetization, $\sups$. This dimensionless parameter is defined as the ratio of the magnetic and the plasma enthalpy densities evaluated for the upstream unreconnected region. Typically hard power laws (i.e., $p\gtrsim -2$) are produced when the magnetization is high (i.e., $\sups\gtrsim 10$) \citep{2014PhRvL.113o5005G, 2014ApJ...783L..21S, 2016ApJ...816L...8W}.

The exact mechanism of particle acceleration and power-law formation in relativistic reconnection has been the topic of extensive research. Possible candidates include direct acceleration in magnetic X-points \citep[e.g.,][]{2001ApJ...562L..63Z, 2011ApJ...737L..40U, 2014ApJ...783L..21S}, Fermi-like acceleration by the motional electric field via the so-called ``slingshot'' mechanism \citep[e.g.,][]{2017ApJ...843...21L, 2019ApJ...879L..23G}, and mergers between large plasmoids \citep[e.g.,][]{2006Natur.443..553D, 2015ApJ...815..101N}. We will further refer to these mechanisms as {\it pre-acceleration} (or {\it primary} acceleration), while their details will remain out of the scope of this paper. 

So far the pre-acceleration stage has been under the spotlight of the community. Instead, our main focus will be the energization process operating on longer timescales after the pre-acceleration stage, which we will refer to as the {\it secondary} acceleration. This secondary process has often been neglected in previous studies because it can only be seen by evolving a large-enough system to long timescales. \citealt{2018MNRAS.481.5687P} (hereafter PS18) performed large 2D simulations, where they demonstrated that a power law is formed at relatively short timescales during the pre-acceleration stage, while particles are slowly energized during the secondary acceleration stage on much longer timescales. In particular, they showed that in late stages of reconnection the characteristic maximum energy of the population of particles increases sublinearly with time, $E_{\rm max} \propto t^{1/2}$. This secondary acceleration, while being slow, may have an imprint on the formation and evolution of the nonthermal tail in the particle spectrum on long timescales, and might be relevant for astrophysical applications. 

In this paper, we expand on the work of PS18 by investigating in detail the secondary particle energization process. In Section~\ref{sec:qualitative}  we discuss qualitatively the structure of the reconnection layer and its dynamics. In Section~\ref{sec:analytics} we introduce our analytical model of the secondary acceleration and the formation and evolution of the nonthermal particle energy spectrum. Our analytical model relies on certain physical assumptions about both the structure of plasmoids and the motion of particles within them. To verify our analytical model, we perform numerical simulations with a setup presented in Section~\ref{sec:setup}. In Sections~\ref{sec:plasmoids} and  \ref{sec:particles_in_plasmoids} we justify the assumptions of our analytical model and empirically demonstrate their validity using results of our simulations. In Section~\ref{sec:discussion} we discuss our results, focusing on their applicability to astrophysical systems. We conclude in Section~\ref{sec:summary} with a summary of the most important findings of this paper.

%
%
%
%

\section{A qualitative overview of the reconnection layer}
\label{sec:qualitative}
\begin{figure*}[htb]
    \centering
    \makebox[2\columnwidth][c]{
        \includegraphics[width=2\columnwidth]{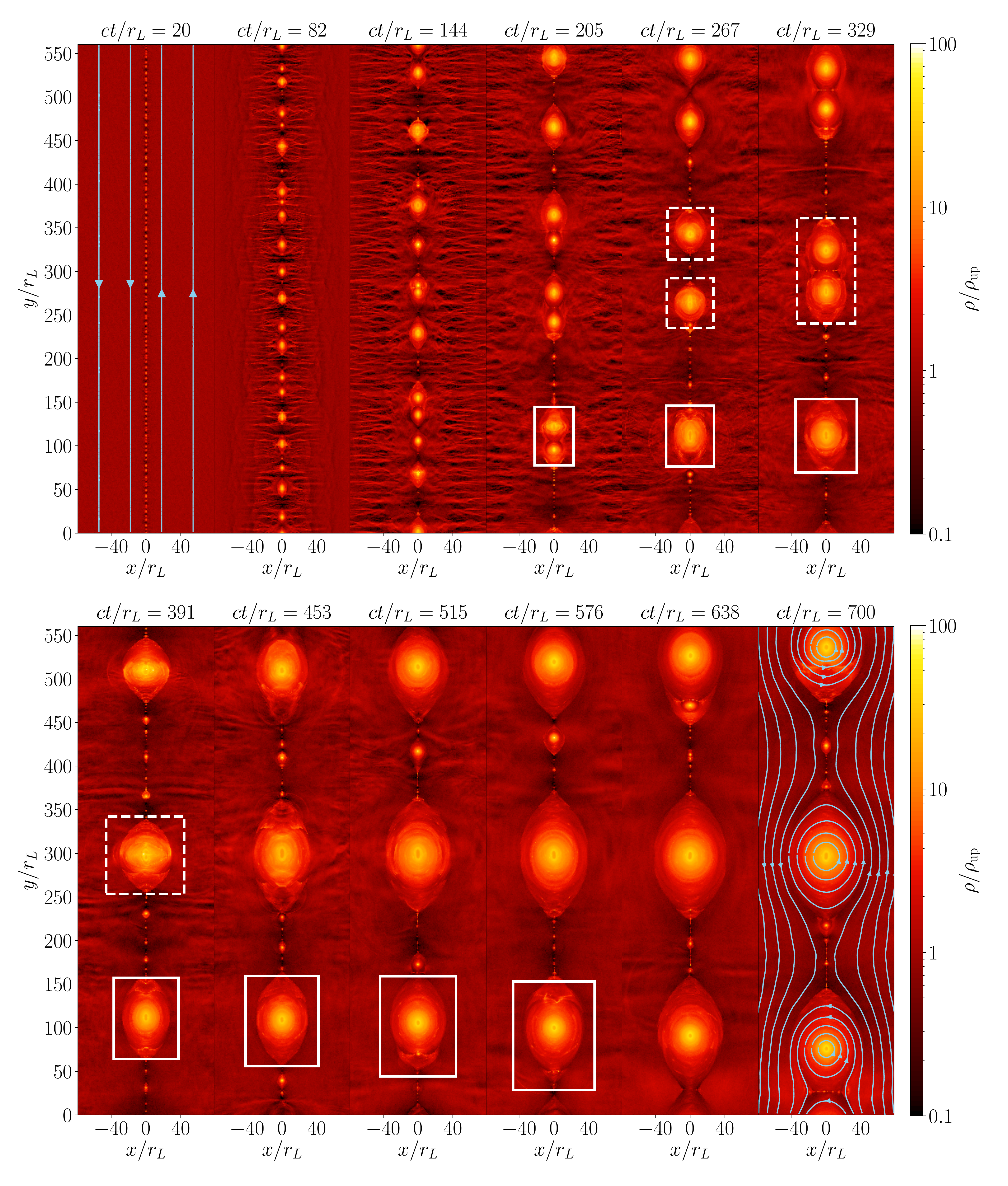}
    }
    \caption{Snapshots showing the temporal evolution of the current sheet from a simulation with the magnetization of the background (upstream) plasma of $\sigma_{\rm up}=100$. Color represents the plasma mass density $\rho$, in units of the mass density in the upstream region $\rho_{\rm up}$, in logarithmic scale (see color bar). We only show the region $|x| / r_L < 60$ to emphasize the small-scale structures in the reconnection layer, while the actual simulation box spans from $-200\, r_L$ to $200 \, r_L$ in the $x$-direction.  The plasmoid used in our subsequent analysis (see Section~\ref{sec:plasmoids}) is highlighted with a solid white rectangle. Dashed white rectangles track the collision of two primary plasmoids (at $ct/r_L\sim322$) from the pre-merger ($ct/r_L=247$) to the post-merger ($ct/r_L=398$) phases. In the first and last panels we also overplot the magnetic field lines for reference. Here we used as our unit of length the Larmor radius $r_L$ of particles with energy $\sigma_{\rm up} m_e c^2$ (for the exact expression, see Equation~\eqref{eq:rL}).}
    \label{fig:evolution}
\end{figure*}

We qualitatively describe the structure and evolution of the reconnection layer in the relativistic regime, setting the stage for the analytical model of particle energization presented in the next section.

Figure~\ref{fig:evolution} shows snapshots of the plasma density structure from a 2D simulation of reconnection in pair plasma. The simulation is initialized with a cold background (upstream) plasma and a hot dense current sheet in the middle ($x=0$); the magnetic field in the $y$-direction changes its sign at $x=0$ (for a detailed description of the simulation setup, see Section~\ref{sec:setup}). At early times the current sheet ``breaks'' in several locations as a result of the tearing instability \citep{1977PhFl...20.1341D, 2005PhRvL..95i5001Z, 2005ApJ...618L.111Z}, which in our simulations develops from numerical noise. Tearing of the initial current sheet leads to the formation of a series of primary magnetic islands, or {\it plasmoids}. These are separated by X-points, i.e., locations where the magnetic field vanishes, introducing a nonideal electric field. Secondary current sheets are formed in between primary plasmoids, and over time they also become unstable, leading to the formation of {\it secondary plasmoids} \citep{2006GeoRL..3313105D, 2010PhRvL.105w5002U, 2016PhRvL.116j5003U}. Although primary and secondary plasmoids evolve in a similar way, they have different internal structures. More specifically, primary plasmoids have an unmagnetized core with plasma from the initial current sheet, while secondary plasmoids form from the secondary current sheets that have been enrinched with magnetized upstream plasma. Henceforth, we focus on the evolution and structure of primary plasmoids, as they will contain the highest-energy particles in our simulations (see also PS18).

Plasmoids grow in size as they continuously accrete plasma and magnetic flux from the upstream region and flows of reconnected plasma along the current sheet (see, e.g., plasmoid highlighted with a solid white rectangle in Figure~\ref{fig:evolution}). While plasmoids grow, their interiors compress over time, as injected particles and magnetic flux are advected inward toward the plasmoid center. Plasmoids can also collide and merge with one another to form bigger islands. In addition to ``minor'' mergers between plasmoids of unequal sizes, a plasmoid can occasionally undergo a ``major'' merger when colliding with a plasmoid of similar size (or equivalently similar mass), as illustrated in Figure~\ref{fig:evolution} with a dashed white rectangle. In this paper, we will focus on periods between major mergers during which the properties of plasmoids (e.g., size, magnetic flux, and mass) evolve adiabatically slowly, i.e., at a rate dictated by the plasma inflow into the current sheet (this will be demonstrated in detail in Section~\ref{sec:particles_in_plasmoids}).

In general, the energy spectrum of particles injected into isolated plasmoids comprises of two main populations: cold particles (i.e., directly accreted from the upstream region), and nonthermal particles pre-accelerated in the regions of reconnected plasma (e.g., X-points, relativistic outflows along the sheet, smaller plasmoids). The exact shape of the injection spectrum will depend on the relative contribution of the two particle populations and its evolution with time. In our simulations, we typically find that the particle injection spectrum into isolated plasmoids can be phenomenologically described by a power law extending in energy up to a few times $\sigma_{\rm up} m_e c^2$ (for details, see Section \ref{sec:particles_in_plasmoids}).

Given this pre-accelerated energy spectrum, we aim to study the long-term energy evolution of particles upon their injection into magnetic islands, a process we refer to as the {\it secondary acceleration}.

\section{An analytical model for particle energization in plasmoids}
\label{sec:analytics}
In order to highlight the main mechanisms at work, we build an analytical model for the long-term particle energization within a constantly compressing plasmoid. 

Motivated by our simulation results (see Sections \ref{sec:plasmoids} and \ref{sec:particles_in_plasmoids}) we assume that the magnetic field lines in the plasmoid interior can be described as concentric rings (see also last panel of Figure~\ref{fig:evolution}). The radius of each ring is decreasing with time, while its magnetic field strength is increasing as a result of plasmoid compression. Particles within plasmoids are typically strongly magnetized (i.e., their gyroradius is much smaller than the plasmoid size), and their motion is confined to the concentric shrinking magnetic rings. Particles injected into the plasmoid roughly at the same time are tied to a single ring and experience an increasing magnetic field strength in time.

We model a compressing plasmoid in the reconnection layer as a confined region wherein charged particles are constantly injected. The volume of this region is permeated by a uniform magnetic field of increasing strength, $B(t)$, due to compression. The change in the magnetic field strength is assumed to be slow compared to the gyration period of particles. \retwo{The first adiabatic invariant for particles is therefore conserved, $\mu \propto u_{\perp}^2 / B \propto \mathrm{const}$, where $u_{\perp}$ is the particle four-velocity in the direction perpendicular to the magnetic field. Here, for simplicity, the particle motion is considered to be confined in the plane perpendicular to the magnetic field, so that $u = u_{\perp}$, and we consider only the evolution of $f(t;u_\perp)$. This toy model is sufficient for explaining the trends found in our simulations, because the distribution cutoff energy and the high-energy tail are largely dictated by $u_{\perp}$. In the further discussion we also consider only the high-energy tail of the distribution function, where for particles $u\approx \gamma \gg 1$.

The evolution of the Lorentz factor of a single particle at the high-energy end of the distribution is then described by
\begin{equation}
    \label{eq:gammadot}
    \dot{\mu}=0~~~\rightarrow~~~\dot{u}_\perp\approx\frac{u_\perp}{2}\frac{\dot{B}}{B}~~~\rightarrow~~~\dot{\gamma}\approx\frac{\gamma}{2}\frac{\dot{B}}{B}.
\end{equation}
}
For a power-law scaling with time, i.e., $B\propto t^{\alpha}$, solution of Equation (\ref{eq:gammadot}) yields $\gamma\propto t^{\alpha/2}$ in the limit of $\gamma\gg 1$. In the case of linear growth of the magnetic field strength with time (i.e., $\alpha=1$), the particle energy will scale as $\propto t^{1/2}$. This is in agreement with the findings of PS18 about the growth of the maximum particle energy. Henceforth, we will assume for simplicity that $B(t)=B_0 \left(t/t_0\right)$. Equation (\ref{eq:gammadot}) then reads $\dot{\gamma}=\gamma/2t$.

Let us now consider the evolution of the distribution function, $f(t,\gamma)$, of the particle population contained in the volume. This evolution can be described by the following equation:
\begin{equation}
    \label{eq:fevolution}
    \frac{\partial f}{\partial t} + \frac{\partial}{\partial \gamma}\left(f\dot{\gamma}\right)=S(t,\gamma),
\end{equation}
where $S(t,\gamma)$ is a source term describing particle injection into the fixed volume. Because particles are confined within the volume (as it happens in plasmoids), there is no escape term on the left-hand side of the equation. Notice that in Equation~\eqref{eq:fevolution} for simplicity, it is assumed that particles injected at time $t_1$ will start experiencing a background magnetic field of strength $B(t_1)$ (because the energization rate $\dot{\gamma}(t)$ is common for all the injected particles). In reality, for plasmoids particles would have started from the upstream field $B(t_0)$ regardless of when they are injected. This, however, does not affect the final outcome, because the highest-energy part of the plasmoid spectrum is populated by the oldest particles, i.e., those that have been injected first in the plasmoid.\footnote{We have carried out synthetic particle simulations with individual particles being injected and getting energized according to Equation~\eqref{eq:gammadot}, i.e., always starting with $B(t_0)$. Results of these runs show that Equation~\eqref{eq:fevolution} approximates well the high-energy part of the resulting distribution function.}

\retwo{The approach of using equations similar to Equation~\eqref{eq:fevolution} to describe the evolution of the power-law distribution of particles during magnetic reconnection is not novel. Similar approaches have been used earlier to study the primary acceleration and the emerging distribution of particles assuming first-order Fermi energization mechanism to estimate $\dot{\gamma}$ and $f(t;\gamma)$ \citep[see, e.g.,][]{2012MNRAS.422.2474D, 2014PhRvL.113o5005G, 2017PhPl...24f2906M, 2019ApJ...879L..23G}. In our case, however, a simplified approach is employed, where the energization term directly follows from the magnetic moment conservation \eqref{eq:gammadot}, and since particles are confined within the plasmoids, there is no escape term on the right-hand side. Moreover, in our model the magnetic field strength, $B$, and the plasma density, $\rho$,  are coupled via the MHD force balance condition and the equation of state (EOS) within plasmoids (for details, see Section~\ref{sec:plasmoids}).}

Assuming that at $t=t_0$ the volume is empty (i.e., $f(t_0,\gamma)=0$), and using the equation $\dot{\gamma}=\gamma/2t$, we obtain the general solution of Equation~\eqref{eq:fevolution}, which reads
\begin{equation}
    \label{eq:fsolution}
    f(t,\gamma)=\frac{2t}{\gamma^3}\int_{\gamma\sqrt{t_0/t}}^{\gamma}\xi^2 S\left(\xi\frac{\sqrt{t}}{\gamma},\xi\right)d\xi.
\end{equation}

\begin{figure*}[htb]
    \centering
    \includegraphics[width=2\columnwidth]{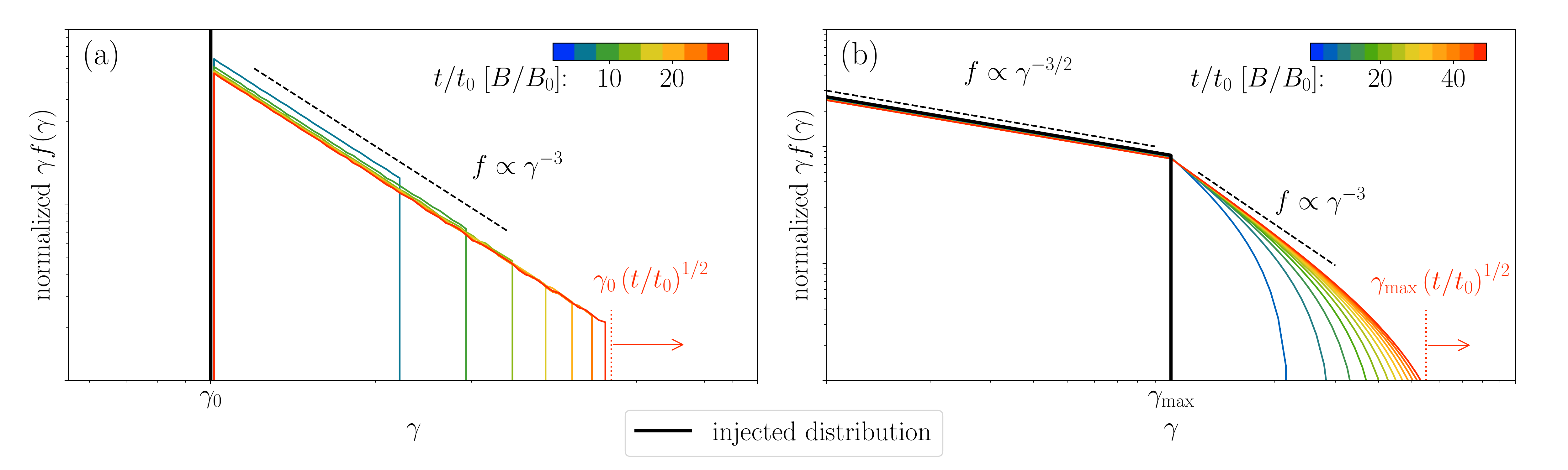}
    \caption{Temporal evolution of the particle distribution function, $f(t, \gamma)$ (see color bar), as obtained by numerically solving Equation~\eqref{eq:fsolution} with monoenergetic (panel (a)) or power law with $\gamma_{\min} \ll \gamma_{\max}$ (panel (b)) distribution functions at injection, $S$ (shown in black). In both panels, the magnetic field grows linearly with time, while the particle injection rate is assumed to be constant. All distribution functions are normalized, so that $\int f(\gamma)d\gamma = 1$. }
    \label{fig:theory}
\end{figure*}

In Figure~\ref{fig:theory} we plot Equation~\eqref{eq:fsolution}, for two different choices for the source term, $S(t,\gamma)$. Curves with different colors correspond to different times, $t/t_0$ (and equivalently to different magnetic field strengths, $B/B_0\equiv t/t_0$), as indicated by the inset color bars. In both panels, black solid lines indicate the distribution function of injected particles.

In the simplest scenario, when particles are injected into the volume at a constant rate, $\dot{n}=\mathrm{const}$, and with the same energy $\gamma_0$ (i.e., $S(t, \gamma)=\dot{n}\delta(\gamma - \gamma_0)$), a power-law distribution function  will develop over time, namely, $f(\gamma)\propto \gamma^{p}$ with $p=-3$, extending to an evolving high-energy cutoff $\gamma_{\rm cut}=\gamma_0\left(t/t_0\right)^{1/2}$ (see Figure~\ref{fig:theory}(a)). 


We consider next a scenario where particles are injected into the compressing volume with a power-law injection spectrum at a constant rate (i.e., $S(t, \gamma)=\dot{n}\gamma^{s} H(\gamma - \gamma_{\min})H(\gamma_{\max}-\gamma)$, with $\gamma_{\rm max}\gg \gamma_{\rm min}$, where $H(\gamma)$ is the Heaviside function). This scenario is inspired by our simulation results, which will be described in detail in Section~\ref{sec:particles_in_plasmoids}, where particles are injected into plasmoids already pre-accelerated. The evolution of the particle distribution function in this case is illustrated in Figure~\ref{fig:theory}(b), for $s=-3/2$. The shape of the distribution in the range $\gamma < \gamma_{\rm max}$ resembles the injected power law. Thus, a power-law distribution with a sharp cutoff at $\gamma_{\max}$ upon injection will be transformed into a broken power law with a sharp break at $\gamma_{\max}$, as shown in Figure~\ref{fig:theory}(b). The break indicates the transition from the injected power law to a power law with an asymptotic slope $p=-3$, as expected by monoenergetic injection of particles at $\gamma_{\max}$. If the high-energy cutoff of the injected spectrum is not sharp, then a smooth transition between the two power-law segments is expected instead of the sharp break at $\gamma=\gamma_{\max}$ shown in Figure~\ref{fig:theory}(b).

\retwo{In this simplified model particles were confined to move in the direction perpendicular to the magnetic field, i.e., Equation~\eqref{eq:fsolution} describes the evolution of the $f(t;\gamma_{\perp})$. In our model, as we argue in Section~\ref{sec:conserv_ad_inv}, energizations in parallel and perpendicular direction are disentangled. At the same time, for the highest-energy particles $\gamma_{\perp} \gtrsim 2\gamma_\parallel$, meaning that the cutoff and the power-law slope are well described by the evolution of $f(t;\gamma_{\perp})$.

The results will be slightly modified for the case when the energy gain in the parallel and perpendicular directions is entangled via the enforced isotropy condition.\footnote{A discussion of possible isotropization mechanisms as well as the associated timescales, can be found in Section~\ref{sec:anisotropy}.} While Equation~\eqref{eq:fevolution} would still be applicable, the energization term, $\dot{\gamma}$, would become $\dot{\gamma} = \dot{\gamma}_{\perp} (\gamma_\perp/\gamma)+\dot{\gamma}_{\parallel} (\gamma_\parallel/\gamma)$. In the extreme case where the distribution isotropy is enforced on timescales shorter than the acceleration timescale, i.e., $\langle\gamma_\parallel\rangle/\langle\gamma_\perp\rangle=1/2$ at all times, we obtain $\dot {\gamma} = 3\sqrt{5}\dot{\gamma}_\perp/4 = 3\gamma / 4 t$, resulting in a somewhat faster acceleration rate, $\gamma_{\rm cut}\propto t^{3/4}$. However, the slope of the power-law tail will remain unchanged. 
}

\reone{An important remark is that the $p=-3$ power law is not universal and strictly relies on the magnetic field compression rate, $B\propto t^{\alpha}$, and the parameter $\alpha$ (which for $p=-3$ is $1$). This parameter, as will be shown in consequent sections, depends on the plasmoid growth rate and does not vary significantly (see Section~\ref{sec:plasmoids}). So in general it is safe to assume that for realistic parameters the emerging power law will be very close to $p=-3$. }

Let us briefly recap the main results of our analytical model, which relies on the magnetic field compression and conservation of the first adiabatic invariant of particles.
\begin{itemize}
    \item Particles injected at a constant rate and with the same (relativistic) energy into an isolated volume permeated by a magnetic field whose strength is increasing linearly with time will obtain over time a power-law distribution function with slope $p=-3$ extending up to a high-energy cutoff evolving as $\propto t^{1/2}$.
    \item Particles injected at a constant rate with a power-law distribution function into the same volume will obtain over time a broken power-law distribution function with a break at the high-energy cutoff of the injection spectrum. The shape of the distribution below the break is the same as upon injection, while the  power-law segment above the break has a slope $p=-3$ and is followed by a high-energy cutoff evolving as $\propto t^{1/2}$. 
\end{itemize}

In subsequent sections we will address several assumptions that are used in the analytical model and may appear ad hoc. In Section~\ref{sec:plasmoids} we present a theoretical model for the plasmoid interior structure that is developed based on the findings of our numerical simulations, whose setup is described in Section~\ref{sec:setup}. We later combine the model for the plasmoid structure with the dynamics of particles within plasmoids to justify our analytical model for the evolution of the particle energy spectrum. In Section~\ref{sec:particles_in_plasmoids} we discuss the temporal evolution of particles injected into the plasmoid and directly compare the analytical predictions with our simulations.

%
%
%
%

\section{Simulation setup} \label{sec:setup}
We use the electromagnetic relativistic particle-in-cell code \texttt{TRISTAN-MP v2},\footnote{\url{https://ntoles.github.io/tristan-wiki/}} which is a multispecies extension of the original \texttt{TRISTAN-MP} code \citep{2005AIPC..801..345S}. 
We perform 2D simulations of reconnection in electron--positron (pair) plasmas with zero guide field. We initialize the reconnection layer (along the $y$-direction) as a Harris sheet with length $L$. The magnetic field 
\begin{equation}
    \bm{B} = B_{\rm up} \tanh(x/\Delta) \hat{\bm{y}},
\end{equation} 
reverses at $x=0$ over a thickness $\Delta$. We choose the latter to be small enough so as to make the current sheet tearing-unstable on short timescales. For that we typically use $\Delta\approx 5\left(c/\omega_{\rm p}\right)_{\rm up}$ and $L\approx 5000\left(c/\omega_{\rm p}\right)_{\rm up}$, where $\left(c/\omega_{\rm p}\right)_{\rm up} \equiv \sqrt{m_e c^2/ 4\pi n_{\rm up} e^2}$ is the skin depth of the cold upstream plasma, which we resolve with five simulation cells, and $n_{\rm up}$ is the number density of background electrons (or positrons). Thus, even if we do not perturb the initial current sheet, it will ``break up'' starting from numerical noise with a subsequent development of the plasmoid instability. We use periodic boundaries in the $y$-direction (which is parallel to the current sheet), while in the other direction our boundaries are open with constant injection of plasma and magnetic field (for details, see \citealt{2014ApJ...783L..21S}). \retwo{The energy in our simulations is not conserved to machine precision owing to the explicit nature of the numerical scheme and finite number of particles per skin depth. We thus employ eight Gaussian filter passes on deposited currents, which keeps the energy nonconservation well below the $1\%$ level. Note also that inside plasmoids the number of particles per skin depth is $\mathcal{O}(10^2)$, which further decreases the numerical noise in these regions of interest.}

Upon initialization, the term $\nabla\times\bm{B}$ is balanced by the out-of-plane current, $j_z$. The magnetic pressure outside the current sheet is balanced by the particle pressure in the initial current sheet, which is provided by a hot plasma with three times higher number density compared to the number density of particles outside the layer. Because the properties of these initially hot particles in the current sheet depend on initial conditions, we exclude them from further analysis.
  
The plasma outside the layer (upstream) is cold, with a small thermal spread upon initialization ($k T_{\rm up}/m_e c^2 \equiv \Theta_e =10^{-4}$). The key parameter that characterizes the overall dynamics of the system is the magnetization of the upstream plasma, $\sigma_{\rm up}$. This quantity is a dimensionless measure of the magnetic energy available per particle and can be written as
\begin{equation}
    \label{eq:sigmaups}
    \sigma_{\rm up} = \frac{B_{\rm up}^2}{4\pi h},
\end{equation}
where $h$ is the plasma enthalpy density of the upstream plasma, including the contribution of its rest-mass energy density, i.e., 
\begin{equation}
    \label{eq:enthalpy}
    h = \rho_{\rm up}c^2\left(1 + \frac{\Gamma}{\Gamma-1} \Theta_e\right),
\end{equation} 
with $\Gamma$ being the adiabatic index of the plasma and $\rho_{\rm up}=n_{\rm up} m_e$. In the case of cold upstream plasma ($\Theta_e \ll 1$), as considered here, the enthalpy density is simply given by the rest-mass energy  density of the plasma, and the magnetization simplifies to $\sigma_{\rm up} = B_{\rm up}^2/4\pi \rho_{\rm up} c^2$. In this paper, we study reconnection in the relativistic regime (i.e., $\sups\gg 1$) and show results from two large-scale simulations with $\sups=10$ and $100$.

In general, the Larmor radius of an electron (or positron) in the upstream magnetic field, $B_{\rm up}$, can be written as 
\begin{equation}
    \tilde{r}_{L} = \gamma\beta \frac{m_e c^2}{|e| B_{\rm up}} = 
    \gamma\beta\frac{\left(c/\omega_{\rm p}\right)_{\rm up}}{\sqrt{\sups}}.
\end{equation}
where $\gamma$ and $\beta$ are the particle's Lorentz factor and three-velocity (in units of $c$), respectively.\footnote{Here the motion is assumed to be confined in the direction perpendicular to the magnetic field.} The Larmor radius, $r_L$, of particles with $\gamma=\sigma_{\rm up}\gg 1$, $\beta\approx 1$, which roughly corresponds to the energy gain assuming that particles tap the whole dissipated magnetic field energy, is
 \begin{equation}
 \label{eq:rL}
    r_{L}=\sqrt{\sigma_{\rm up}}\left(\frac{c}{\omega_{\rm p}}\right)_{\rm up}.
\end{equation}
Henceforth, we adopt $r_{L}$ as our length unit, and we quote times normalized to $r_L/c$. 

%
%
%
%

\section{Structure of plasmoids}
\label{sec:plasmoids}

\begin{figure}[htb]
    \centering
    \includegraphics[width=\columnwidth]{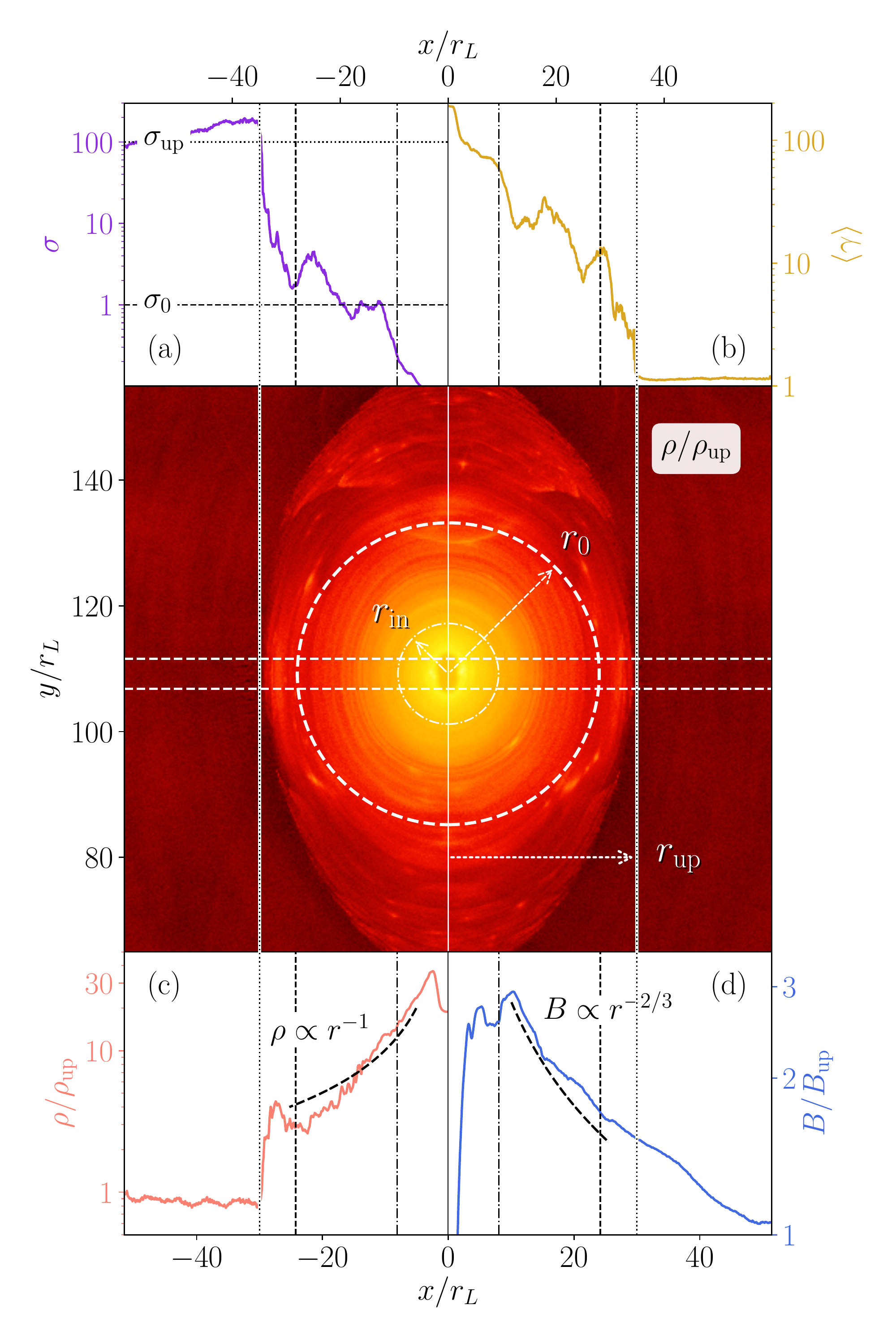}
    \caption{Close-up view of a representative isolated primary plasmoid from the simulation with $\sups=100$ at time $ct/r_L=391$; this plasmoid is indicated in Figure~\ref{fig:evolution} by a white solid rectangle. Color in the middle panel represents the plasma density $\rho$, in units of the upstream plasma density $\rho_{\rm up}$, in logarithmic scale (same color-coding as in Figure~\ref{fig:evolution}). Three characteristic radii are also marked on the plot: $\rups$ indicates the boundary of the {\it plasmoid corona} (i.e., where the plasmoid ends and the upstream begins); $r_0$ shows the outer boundary of the {\it plasmoid shell}, where the local force balance condition is satisfied; and $\rins$ indicates the plasmoid core that contains hot unmagnetized plasma from the initial current sheet. Four peripheral panels show the radial profiles of (a) the magnetization, (b) mean Lorentz factor, (c) plasma density in units of its upstream value, and (d) magnetic field in units of its upstream value, computed along a transverse stripe passing through the plasmoid center (see horizontal dashed white lines in middle panel). In all peripheral panels, vertical lines indicate the characteristic radii marked in the middle plot. Two horizontal lines in the top left panel indicate the upstream magnetization, $\sups \gg 1$, and the effective magnetization of the plasmoid shell, $\sigma_0 \approx 1$ (see Appendix \ref{appendix:plasm}). All plots share the same $x$-axes.}
    \label{fig:plasm_example}
\end{figure}

As we have postulated in Section~\ref{sec:analytics}, plasmoids can be thought of as compressing regions with a constantly amplifying magnetic field. They continuously accrete particles from the upstream plasma and the reconnected plasma outflows. In this section, we will study the structure of plasmoids and the plasmoid compression rate and explore the factors that determine these plasmoid properties. 

Let us take a close look at the structure of a typical isolated plasmoid in the reconnection layer. Figure~\ref{fig:plasm_example} (middle panel) shows a close-up view of the plasma density structure in a typical isolated primary plasmoid, also highlighted with a solid rectangle in Figure~\ref{fig:evolution} at time $ct/r_L = 391$. The four peripheral panels in the figure show 1D profiles of the magnetization\footnote{
The magnetization, $\sigma$, is computed using the local {magnetic field strength and} enthalpy density, $\sigma = B^2 / 4\pi h$, where  $h$ is defined in Equation~\eqref{eq:enthalpy} with $\Gamma\sim 4/3$ and $\Theta_e\equiv P_e/\rho_ec^2 \sim \langle\gamma\beta^2\rangle$/3.}
($\sigma$, panel (a)), mean particle Lorentz factor ($\langle\gamma\rangle$, panel (b)), mass density ($\rho$, panel (c)), and magnetic field strength ($B$, panel (d)), computed along a transverse stripe of width $5r_L$ passing through the plasmoid center (white dash-stroked stripe in middle panel). 
All panels share common $x$-axes. 

As we get closer to the plasmoid center, we can see how magnetization, $\sigma$, drops from the upstream value, $\sups$, to about $\sigma_0\sim 1$, which implies equipartition between the plasma and the magnetic field energy densities. At the same time, the plasma gets hotter towards the center (see panel (b) for $\langle\gamma\rangle$), which suggests that within the plasmoid the plasma has already been heated by magnetic energy dissipation. Both the plasma density and magnetic field strength increase compared to the upstream values as power laws of the distance from the plasmoid center (lower panels).

In primary plasmoids we can identify three regions of interest that we describe below. The central part of the plasmoid ($r < \rins$), the {\it plasmoid core}, contains typically hot unmagnetized plasma used to initialize the current sheet (see Section \ref{sec:setup}). Because the core bears the memory of our initial conditions, it is excluded from all further analysis. The inner part of the plasmoid beyond the core (i.e., $\rins < r < r_0$), which we label as the {\it plasmoid shell}, is almost circular, and its structure is determined solely by the local force balance condition, which we will discuss later in this section. Henceforth, we use the subscript ``$0$'' to indicate physical quantities computed at the boundary of the plasmoid shell. Finally, the outer part of the plasmoid ($r_0 < r < \rups$),\footnote{$\rups$ denotes the characteristic size of the plasmoid. The algorithm used to determine the plasmoid boundaries is based on the magnetic vector potential and is described in Appendix~\ref{appendix:plasmbound}.} which we call the {\it plasmoid corona}, is elongated along the current sheet. The corona can be thought of as a transitional layer between the inner plasmoid region and the upstream plasma where the magnetization changes rapidly (see, e.g., Figure~\ref{fig:plasm_example}(a)). The coronal dynamics and structure are dictated by the plasma inflow and the time-varying properties of the upstream and current sheet. In what follows, we focus on the structure of the plasmoid shell.

Motivated by the power-law radial profiles of the magnetic field and density in the plasmoid shell (see Figure~\ref{fig:plasm_example}(c) and \ref{fig:plasm_example}(d)), we assume that they can both be expressed as functions of radius from the plasmoid center, $r$, and time, $t$, in the following form:
\begin{equation}
    \label{eq:radialstruct}
    B(r,t)=B_0\left(\frac{r}{r_0(t)}\right)^{-\zeta},~\rho(r,t)=\rho_0\left(\frac{r}{r_0(t)}\right)^{-\xi},
\end{equation}
where $\zeta, \xi \ge 0$, and $B_0\equiv B(r_0(t), t)$, $\rho_0\equiv \rho(r_0(t), t)$ are the time-independent boundary values of the magnetic field and density, respectively. The characteristic size of the plasmoid shell, which is proportional to the plasmoid size at all times (i.e., $r_0(t)\propto \rups(t)$), can be written as
\begin{equation}
    \label{eq:inflationrate}
    r_0(t)\propto t^\kappa,
\end{equation}
where $\kappa\ge 0$. 
Thus, at any fixed radius in the plasmoid shell, the temporal dependence of the magnetic field and plasma density can be written as $B\propto t^{\zeta\kappa}$ and $\rho\propto t^{\xi\kappa}$.

The exact value of $\kappa$ is determined by the large-scale reconnection process. By studying the growth of sufficiently large and slowly moving isolated plasmoids, like the one marked in Figure~\ref{fig:evolution}, we find that
\begin{equation}
\label{eq:kappa}
    \kappa\approx 1/2 - 3/4
\end{equation}
for both $\sups = 10$ and $\sups = 100$ simulations. The exact value of the index $\kappa$ may also depend on the numerical setup and, more specifically, on the boundary conditions used. For example, \cite{2016MNRAS.462...48S} observed $\kappa\approx 1$ 
in their 2D simulations of reconnection with outflow boundary conditions in the
$y$-direction (as opposed to the periodic boundary conditions used in our simulations).


The power-law indices $\zeta$ and $\xi$ of the magnetic field and density profiles (see Equation~\eqref{eq:radialstruct}) can be predicted from the MHD force balance equation for the plasmoid shell, $\bm{j}\times\bm{B}=c{\bm \nabla} P$, assuming a polytropic EOS with adiabatic index $\Gamma$. In Appendix~\ref{appendix:plasm} we show that the force balance condition yields
\begin{eqnarray}
    \label{eq:zeta}
    \zeta & = & \frac{\Gamma\sigma_0/2}{\Gamma + \Gamma\sigma_0/2 - 1}, \\
    \xi & = &  \frac{\sigma_0}{\Gamma + \Gamma\sigma_0/2 - 1} \cdot
    \label{eq:xi}
\end{eqnarray}
Here $\sigma_0$ is the effective magnetization inside the plasmoid shell, which is typically of the order of $\sigma_0\sim 1$, as illustrated in Figure~\ref{fig:plasm_example}(a). For both $\sigma_{\rm up} = 10$ and $\sigma_{\rm up} = 100$ simulations, we also find typical values for the adiabatic index $\Gamma=4/3$ (see Appendix~\ref{appendix:plasm} and Figure~\ref{fig:plasm_eos} for details). Substitution of these values into Equations \eqref{eq:zeta} and \eqref{eq:xi} yields
\begin{equation}
    \zeta \approx 2/3,~\text{and}~\xi \approx 1.
\end{equation}
These values are consistent with what we observe in our simulations (see, e.g., bottom panels in Figure~\ref{fig:plasm_example}) and with the results of \cite{2016MNRAS.462...48S}, who reported $\zeta\approx 0.6$ and $\xi \approx 1$  (see, e.g., Appendix A of that reference).

Let us finally estimate the injection rate of particles into the plasmoid shell. At any given time, the total number of particles in the plasmoid shell can be estimated as 
\begin{equation}
    \label{eq:number_vs_time}
    N_0(t)\propto \int_{\rins}^{r_0(t)} \! \! \rho(r, t) rdr \propto r^2_0(t) \propto t^{2\kappa},
\end{equation}
where we used Equation~\eqref{eq:radialstruct} and assumed that $\rins \ll r_0(t)$ and $\xi \ne 2$ (for $\xi=2$, $N_0(t) \propto r^2_0(t) \ln [r_0(t)/\rins]$). The injection rate can be then written as
\begin{equation}
    \label{eq:injrate}
    \dot{N_0}(t)\propto r_0(t)\dot{r}_0(t)\propto t^{2\kappa-1}.
\end{equation}
For $\kappa\approx 1/2$ the injection rate of particles in the plasmoid shell is exactly constant in time, while for $\kappa\approx 3/4$ the rate scales as $t^{1/2}$. Equation~\eqref{eq:number_vs_time} also implies that the mean density inside the plasmoid shell, $\langle\rho\rangle \propto N_0(t)/r^2_0(t)$, is constant (or scales weakly with time), regardless of the exact value of $\kappa$.  

It is worth noting that the results presented in this section do not directly depend on the upstream conditions, such as the upstream magnetization. The reason is that the interior of the plasmoid -- the plasmoid shell -- contains magnetized plasma that has already been ``reprocessed'' by reconnection. The magnetic flux loops in the plasmoid shell do not bear the memory of the conditions in the unreconnected plasma. They are in force balance with the relativistically hot plasma in the plasmoid shell. The radial profile of the magnetic field essentially depends on the plasma EOS. The growth of the magnetic flux in the plasmoid interior, which is adiabatically slow, is dictated by the global reconnection rate. The reconnection rate, which can also be thought of as the inflow velocity from the upstream, is ubiquitous for systems with low enough-resistivity, and for relativistic plasmas it is equal to $v_{\rm in}\sim 0.1c-0.2c$. 

Summarizing, the key result of this section is that the structure and evolution of the plasmoid shell are described by three dimensionless numbers: the power-law indices of magnetic field and plasma density profiles $\zeta$ and $\xi$, respectively, defined in Equation \eqref{eq:radialstruct}, and the plasmoid growth rate $\kappa$ defined in Equation~\eqref{eq:inflationrate}. The first two are set by the force balance in the plasmoid shell and can be obtained assuming a simple polytropic EOS of the relativistically hot plasma in the plasmoid. The third one, however, is determined by the large-scale reconnection process and has to be determined empirically (from the simulations).

%
%
%
%

\section{Evolution of particles in plasmoids}
\label{sec:particles_in_plasmoids}

\begin{figure*}[htb]
    \centering
    \includegraphics[width=2\columnwidth]{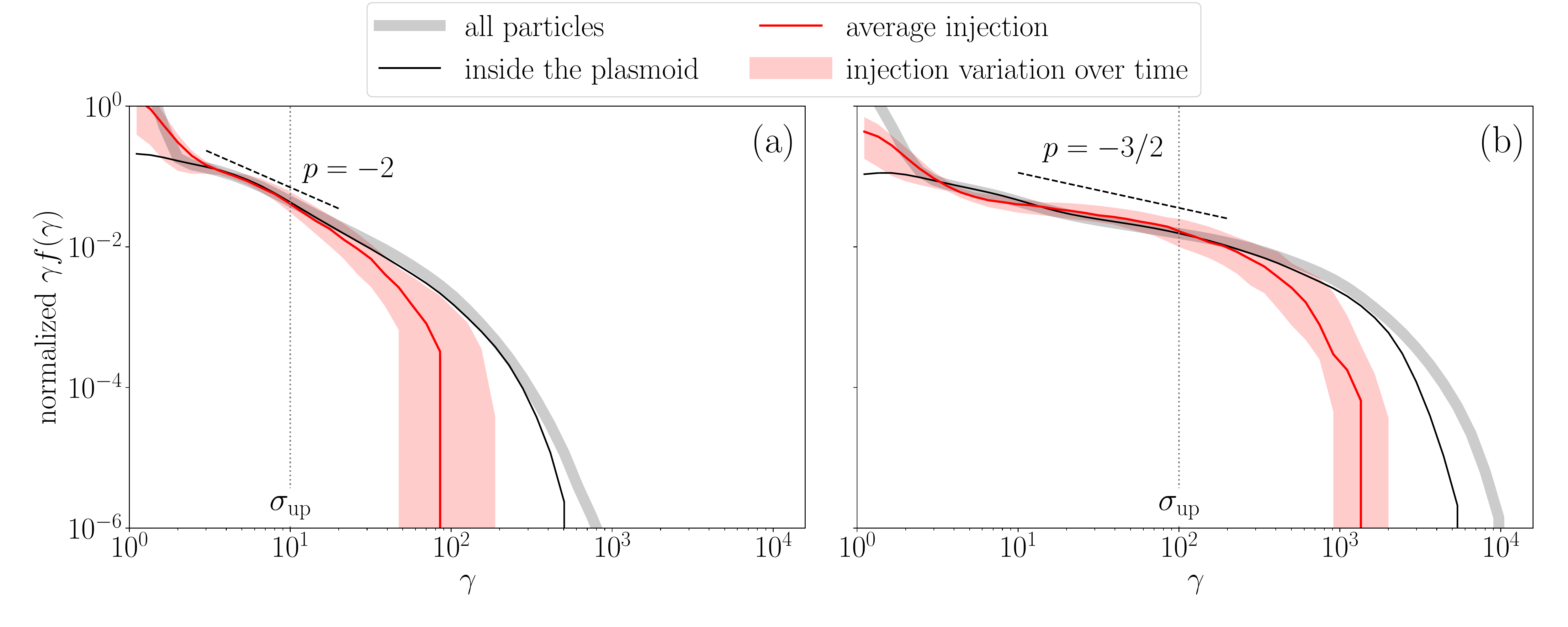}
    \caption{Particle distribution functions, $f(\gamma)$, compensated by $\gamma$, from  the $\sigma_{\rm up}=10$ (panel (a)) and $\sigma_{\rm up}=100$ (panel (b)) simulations. The time-averaged injection spectra are shown with a red line, while the variation of injected distribution over time is illustrated by the transparent red band. For panel (a) the averaging period is $1300<ct/r_L<2000$, and for panel (b) it is $420 < ct/r_L < 620$. Black solid lines show the distribution function of all the particles in the plasmoid, while the thick gray line represents the distribution function of all the particles in the simulation domain; both of these lines are computed at the end of the quoted time period. For comparison purposes, all distribution functions are normalized so that $\int f(\gamma)d\gamma=1$. In these plots it is evident that the distribution function of the plasmoid as a whole (black line) extends farther than the distribution of the injected particles (red line).}
    \label{fig:injspec}
\end{figure*}
In this section, we focus on the evolution of particles upon their injection into plasmoids, while making use of our results about the plasmoid interior structure and its evolution.

There are two main channels for particle injection into plasmoids from the cold upstream region. First, particles can be accreted directly onto a plasmoid as they are carried toward the current sheet by converging magnetic field lines. These particles typically have low energies upon entering a plasmoid (i.e., $\gamma \sim 1$), as they have never interacted with the current sheet before. Alternatively, particles from the upstream region can interact with the current sheet first before entering a plasmoid. In this case, the injected particle population is already pre-accelerated either by the electric field at an X-point \citep{2001ApJ...562L..63Z, 2003ApJ...586...72L, 2008ApJ...682.1436L} or by the motional electric field via the so-called ``slingshot'' Fermi-like mechanism \citep{2006Natur.443..553D, 2014PhRvL.113o5005G, 2015ApJ...806..167G}. Thus, at any given time the injection spectrum of a plasmoid is expected to be a superposition of the ``cold'' component directly coming from the upstream and a ``hot''  pre-accelerated component inflowing from the current sheet. 

Time-averaged spectra of particles injected\footnote{
We compute these spectra using particles near the boundary of plasmoid; to identify the boundary, we use the method described in Appendix~\ref{appendix:plasmbound}. The injection spectra are averaged over the time span mentioned in the caption of Figure~\ref{fig:injspec}.} into a typical isolated plasmoid from our simulations are shown in Figure~\ref{fig:injspec} (red lines). Panels (a) and (b) show results for $\sups=10$ and $\sups=100$, respectively. The plasmoid, whose spectrum is displayed in panel (b), is also highlighted in Figure~\ref{fig:evolution} (white rectangle). In both panels, the average injection spectrum can be described by a power law (i.e., $f(\gamma)\propto\gamma^{p}$ with $p\sim -2$ for $\sups=10$ and $p\sim -1.5$ for $\sups=100$). This power law typically extends up to Lorentz factors of several $\sups$. We also note that the injection spectrum does not vary much with time, as shown  by the red colored band in Figure~\ref{fig:injspec}. This is true except for times very early in a plasmoid's lifetime, when small variations in the amount of mass accreted via the adjacent current sheets and the upstream plasma can significantly affect the overall spectral shape (not explicitly shown here). 

The spectrum computed using all particles trapped within the plasmoid at a given time (thin black line) does not match the injection spectrum, as shown in Figure~\ref{fig:injspec}. More specifically, the energy spectrum of particles contained in the plasmoid appears to be shifted to higher energies compared to the injected spectrum (compare black and red solid lines). These results suggest the presence of an acceleration mechanism operating within the isolated plasmoids that is responsible for pushing the injected particles to even higher energies than those achieved via other processes prior to injection. 

Identifying the process that energizes particles after their injection into plasmoids is also important for understanding the formation of the particle spectrum from the reconnection layer as a whole. The reason is that, at times when the current sheet is dominated by large plasmoids (see $ct/r_L > 300$ in Figure~\ref{fig:evolution}), the majority of particles (including the most energetic ones) are ultimately trapped inside magnetic islands. This is exemplified in Figure~\ref{fig:injspec}, where the particle spectrum from the whole simulation box (thick gray line) -- normalized to the total number of particles -- is compared against that of a single plasmoid (thin black line). When both spectra are normalized to their total number of particles, as done in this figure, the particle spectra for $\gamma \gg 1$ fall on top of each other, except for the highest-energy part. We note also that some of the freshly injected particles (distribution of which is shown by red color) may have already undergone this secondary energization in smaller plasmoids that merged into the bigger plasmoid under study. This, in part, can explain the variability in the particle injection spectrum.

\subsection{Particle evolution in the plasmoid shell}
\label{sec:part-evol}
\begin{figure*}[htb]
    \centering
    \includegraphics[width=2\columnwidth]{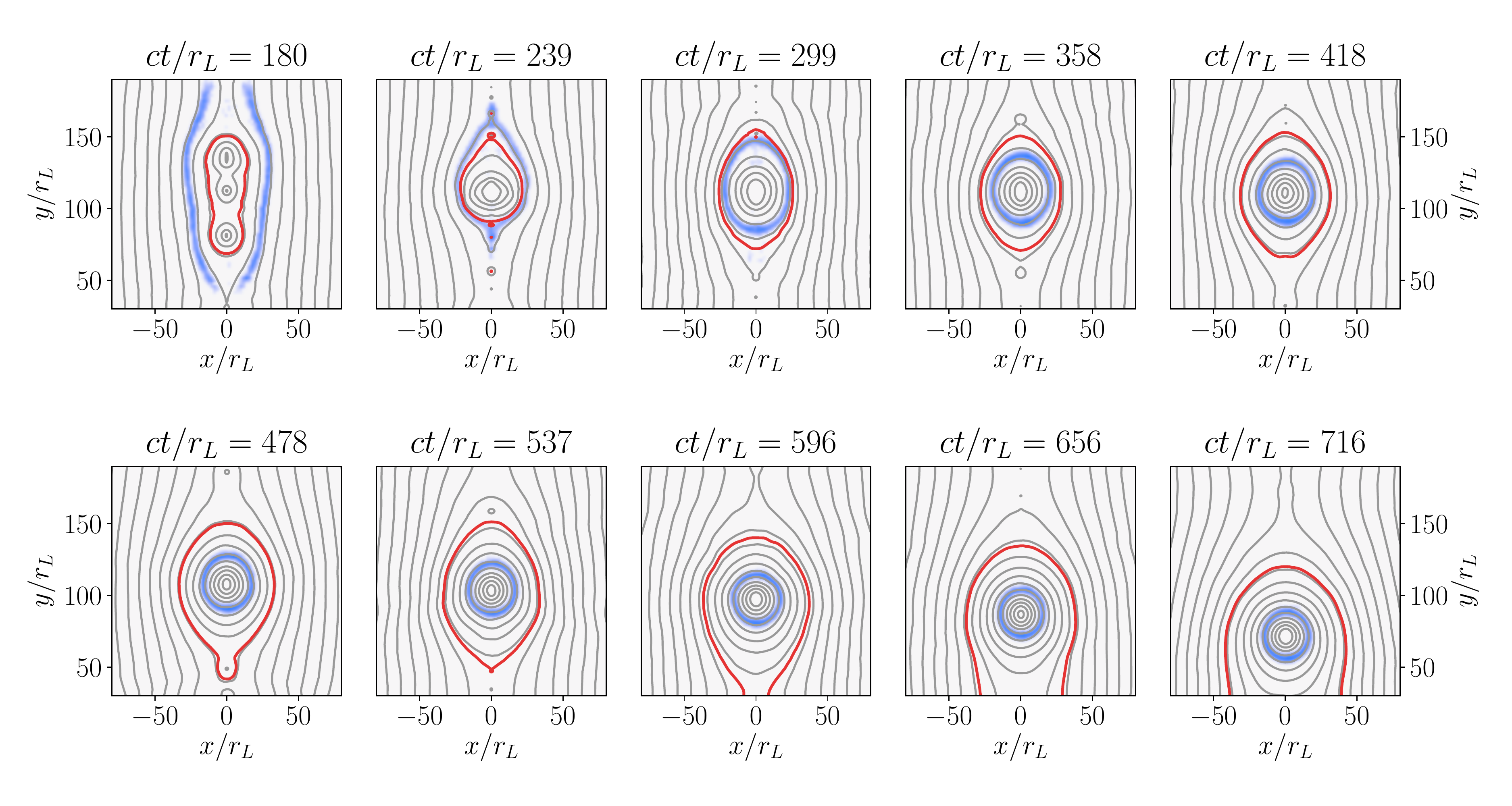}
    \caption{
        Snapshots from the $\sups=100$ simulation focused on the same primary plasmoid highlighted in Figure~\ref{fig:evolution} with solid lines. The plasmoid boundary is highlighted with a solid red line (for details on how we determine the plasmoid boundary, see Appendix~\ref{appendix:plasmbound}). Gray lines are the isocontours of the magnetic vector potential, $A_z$. A population of $\sim 10^4$ particles (shown in blue) is initially frozen to a magnetic field line (see top left panel). The particles enter the plasmoid roughly at the same time ($239 < ct/r_L < 299$) and are later carried toward the center of the plasmoid. An animation showing the evolution of various physical quantities for the same particle population can be found at the following link: \url{https://youtu.be/UJsjoIieLm0}. 
    }
    \label{fig:layerevol}
\end{figure*}
In this section, we track a population of about $10^4$ particles from the $\sups=100$ simulation that enter the isolated primary plasmoid shown in Figure~\ref{fig:evolution} (white rectangle) at roughly the same time (around $ct/r_L\approx 300$) and follow their evolution as they are carried inward to its center. This is illustrated in Figure~\ref{fig:layerevol}, where different panels show different snapshots of the selected plasmoid (its boundary is indicated with a red contour; details about the definition of the plasmoid boundary can be found in Appendix~\ref{appendix:plasmbound}) and the tracked particle population (shown in blue). With the help of these particles, we can not only study the acceleration taking place directly inside the plasmoid but also map the plasmoid structure in Lagrangian terms (i.e., in the frame comoving with the fluid element). Because particles are well magnetized, as their gyroradii are much smaller than the shell size (we will inspect this further in Section~\ref{sec:conserv_ad_inv}), their evolution also tracks the magnetic field line on which they started in the upstream.

In Figure~\ref{fig:layerevol} the particles (shown in blue) are frozen into converging magnetic field lines. At around $ct/r_L\sim 200$ the flux loop reconnects, and some of the particles are exposed to the X-point and are pre-accelerated in the current sheet, forming the injection spectrum shown in red color in Figure~\ref{fig:injspec}(b). Around $ct/r_L\sim 300$ particles cross the plasmoid boundary entering the plasmoid corona but quickly converge into the plasmoid shell, as the flux loop to which they are frozen circularizes (we define the coronae and shells of plasmoids in Section~\ref{sec:plasmoids}).

In the plasmoid shell, particles start their adiabatically slow descent toward the plasmoid core ($ct/r_L > 300$). Plasma in the plasmoid shell is also frozen to the converging concentric magnetic field lines, each of which can be thought of as a circle with a time-varying radius $\Rlg(t)$. Henceforth, calligraphic capital letters will be used to denote variables in Lagrangian terms. The total mass enclosed within a circle of radius $\Rlg$ is constant in time, as particles cannot move across concentric magnetic field lines. This condition yields (for a detailed derivation, see Appendix~\ref{appendix:plasm})
\begin{equation}
    \label{eq:rL_vs_t}
    \Rlg(t)\propto t^{-\kappa\xi/(2-\xi)},
\end{equation}
where $\xi$ and $\kappa$ are defined in Equations~\eqref{eq:radialstruct} and \eqref{eq:inflationrate}, respectively. For $\xi\approx 1$, as found in our simulations (see Section~\ref{sec:plasmoids}), the expression above simplifies to $\Rlg \propto t^{-\kappa}$. The magnetic field strength at the particle location, $\Blg$, can be estimated by substituting $\Rlg(t)$ into Equation~\eqref{eq:radialstruct},
\begin{equation}
    \label{eq:BL_vs_t}
    \Blg(t)\propto t^{2\kappa \zeta/(2-\xi)}\propto t^{4\kappa/3},
\end{equation}
where we assumed $\xi\approx 1$ and $\zeta\approx 2/3$ to derive the second scaling relation in the equation above.

\begin{figure*}[htb]
    \centering
    \includegraphics[width=2\columnwidth]{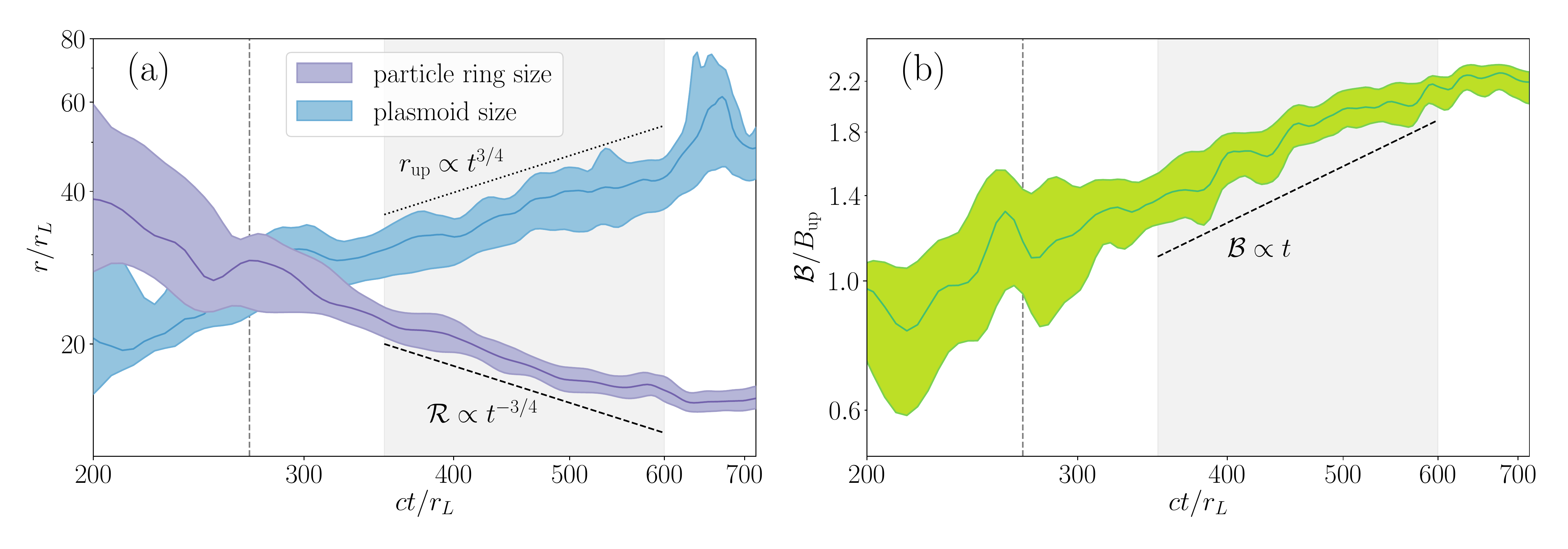}
    \caption{Temporal evolution of the distance from the plasmoid center, $\Rlg$ (panel (a)), and the magnetic field strength at the particle location, $\Blg$ (panel (b)), for the generation of particles shown in Figure~\ref{fig:layerevol} in blue. Colored bands represent the spread in values within the particle population, while the solid lines show the median value. In panel (a) we also show the evolution of the plasmoid size, $\rups$, as a function of time; the spread, in this case, originates from the fact that the outer boundary of this plasmoid is actually elliptical. The vertical dashed line indicates the time when particles enter the plasmoid, and the gray band at $350<ct/r_L<600$ corresponds to the time when particles are within the contracting shell while the plasmoid remains isolated. As particles spiral down toward the center of the plasmoid ($\Rlg(t)$), the plasmoid itself grows ($r_{\rm up}(t)$), and the magnetic field strength that particles experience grows with time ($\Blg(t)$). \reone{At $ct/r_L\sim 600$ the ring of particles reaches the unmagnetized inner core of the plasmoid, where the MHD balance condition we discussed no longer holds, which is why the growth in $\Blg$ halts.}}
    \label{fig:layer_r_b}
\end{figure*}
We then compare the empirical relations for $\Rlg(t)$ and $\Blg(t)$ with the scalings derived directly from our numerical simulations. In Figure~\ref{fig:layer_r_b} we show how the distance from the plasmoid center (purple band in panel (a)) and the magnetic field strength (panel (b)) evolve with time for the same generation of particles shown in Figure~\ref{fig:layerevol}. In both panels, solid lines correspond to the median value of the displayed variable, and the colored band indicates the spread in values within the tracked particle population. As particles move toward the plasmoid center, the corresponding spread in $\Rlg$ and $\Blg$ becomes smaller as the plasmoid shell is circularized. In panel (a), we also plot the radius of the plasmoid boundary, $\rups$, as a function of time (blue band). For this particular plasmoid, we find $\rups\propto t^{3/4}$ or $\kappa\approx 3/4$. As particles are advected by the magnetic loop toward the plasmoid center, their distance decreases with time as $\Rlg\propto t^{-3/4}$, while the magnetic field at the particle location grows roughly linearly with time, $\Blg\propto t$. This is in a good agreement with the analytical scalings of Equations~\eqref{eq:rL_vs_t} and \eqref{eq:BL_vs_t} for $\kappa=3/4$. Notice that the magnetic field strength $\mathcal{B}$ is measured in the lab frame, whereas to compare with our analytical estimations we need to measure it in the frame comoving with the plasmoid (i.e., moving with an $\boldsymbol{E}\times\boldsymbol{B}$ drift velocity). However, since the plasmoid we consider is large and slow, any corrections to our measurements are negligible.

Summarizing, there are two effects acting together to build up the linear increase of the magnetic field strength with time experienced by a particle population after its injection into a plasmoid shell. First, the plasmoid interior gets compressed, and the magnetic field at a fixed distance from its center gets amplified. Second, particles ``sink'' toward the center of the plasmoid, experiencing an increasingly stronger magnetic field. 

\subsection{Conservation of adiabatic invariants}
\label{sec:conserv_ad_inv}
\begin{figure*}[htb]
    \centering
    \includegraphics[width=2\columnwidth]{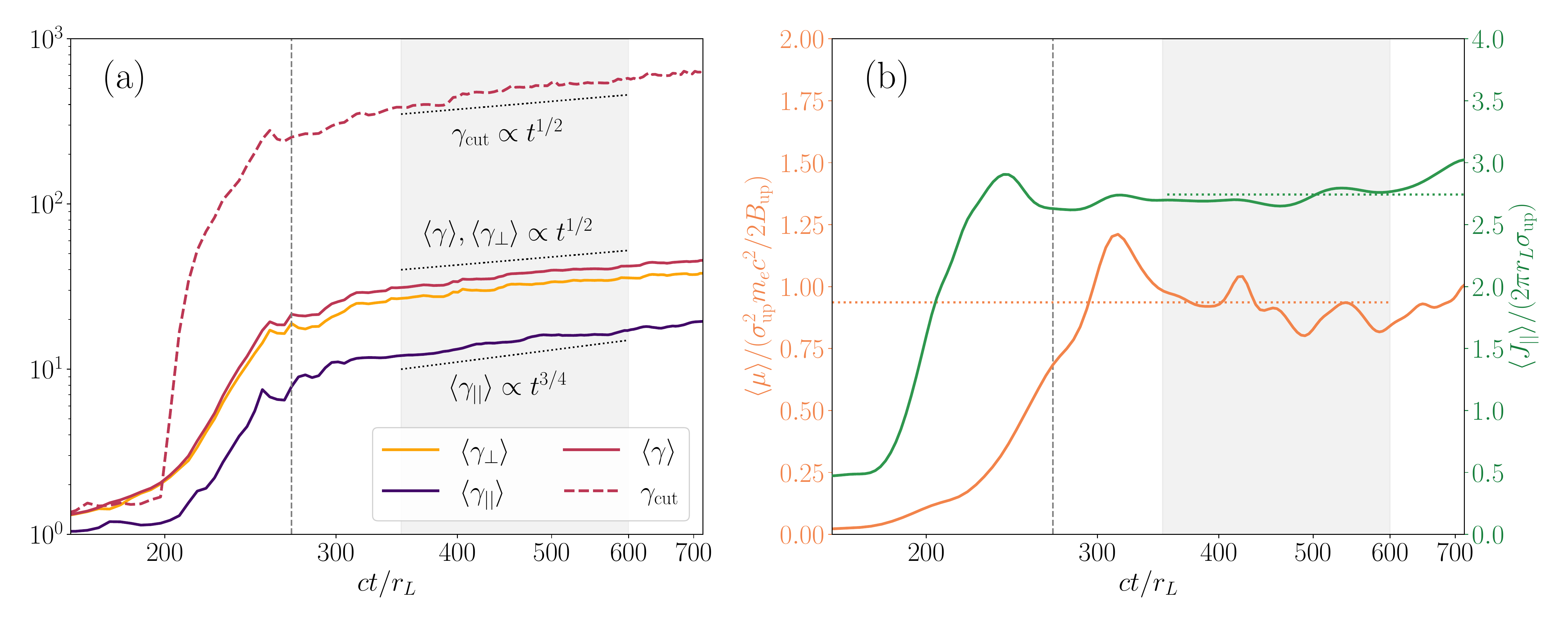}
    \caption{(a)
    Temporal evolution of the particle Lorentz factor $\langle\gamma\rangle$, and the Lorentz factor of the particle motion perpendicular ($\langle\gamma_\perp\rangle$), and parallel ($\langle\gamma_{\parallel}\rangle$) to the magnetic field (see inset legend), averaged over the particle population identified in Figure~\ref{fig:layerevol}. The dashed colored line shows the evolution of the cutoff Lorentz factor of the particle energy spectrum, computed as described in Section~\ref{sec:conserv_ad_inv}. \reone{(b) Temporal evolution of the population averaged adiabatic invariants $\mu$ and $J_\parallel$ (defined in Equations (\ref{eq:mu}) and (\ref{eq:Jpar}), respectively) for the same population of particles as in panel (a). $\langle\mu\rangle$ is normalized to $m_e \sups^2 c^2/2B_{\rm up}$, i.e., the magnetic moment of particles with $\gamma_\perp\sim \sups$ in the upstream field $B_{\rm up}$; $\langle J_\parallel\rangle$, on the other hand, is normalized to $\sups r_L$, i.e., the mirror invariant for a particle with $\gamma_\parallel\sim\sups$ trapped in a circular region of radius $r_L$. These quantities share the same $x$-axis with different $y$-axes on the left and right highlighted with corresponding colors; dotted horizontal lines indicate their average values after $ct/r_L=300$.} 
    In both panels, the gray dashed vertical line shows the moment when particles enter the plasmoid, and the gray band $350<ct/r_L<600$ corresponds to the time when particles are within the contracting shell while the plasmoid remains isolated. As wee see from panel (a), both the parallel and the perpendicular components of particle momenta grow with time. Since particles are well magnetized within the plasmoid shell ($ct/r_L>300$), this energization is caused by the conservation of adiabatic invariants shown in panel (b).}
    \label{fig:layer_gam_mu}
\end{figure*}

We now focus on the energization of particles after they enter into the plasmoid, using the same sample of tracked particles as in the previous section. In Figure~\ref{fig:layer_gam_mu}(a) we plot the evolution of the mean energy of particles, $\langle \gamma\rangle$, as a function of time. For $ct/r_L \gtrsim 290$, i.e., after the particles have sunk into the plasmoid shell (see also Figure~\ref{fig:layerevol}), the growth of the mean energy is sublinear with time. The same applies for the high-energy cutoff of the particle energy spectrum, namely, $\gamma_{\rm cut} \propto t^{1/2}$, in agreement with the results of PS18. The cutoff is found using a similar approach to that in \cite{2015ApJ...809...55B},
\begin{equation}
    \gamma_{\rm cut} = \frac{\int \gamma^{\alpha} f(\gamma)d\gamma}{\int \gamma^{\alpha - 1} f(\gamma)d\gamma},
\end{equation}
where the parameter $\alpha$ is empirically chosen to be $3$. This formula allows one to be agnostic to the exact power-law slope, while roughly estimating the position of the energy cutoff.

In Figure~\ref{fig:layer_gam_mu}(a) we also plot the temporal evolution of the Lorentz factor of the particle motion perpendicular ($\langle\gamma_\perp\rangle$) and parallel ($\langle\gamma_{\parallel}\rangle$) to the magnetic field, averaged over the tracked particle population. We define $\gamma_{\parallel}$ and $\gamma_{\perp}$ of a single particle as
\begin{equation}
    \gamma_{\parallel} = \sqrt{1+u_{\parallel}^2},~\text{and}~\gamma_{\perp} = \sqrt{1+u_{\perp}^2},
\end{equation}
where $u_{\parallel}$ and $u_\perp$ are the parallel and perpendicular components of the particle's dimensionless 4-velocity with respect to the local magnetic field. As in PS18, we find that $\langle \gamma\rangle \approx \langle\gamma_{\perp}\rangle \sim 2 \langle\gamma_{\parallel}\rangle$. This suggests that inside the plasmoid shell the pressure\footnote{We define pressure components as the flux of the corresponding momentum components, $P_i\propto \langle\gamma_i\beta_i^2\rangle \approx \langle\gamma_i\rangle$, because $\beta_i\approx 1$.} is almost isotropic, namely $P_{\perp}\approx \rho c^2\langle\gamma_{\perp}\rangle \approx 2 \rho c^2 \langle\gamma_{\parallel}\rangle \approx 2P_{\parallel}$.

PS18 proposed that the conservation of the first adiabatic invariant, together with the magnetic field amplification due to plasmoid compression, is responsible for the slow and steady energy increase of particles contained within plasmoids. Our results confirm this physical interpretation, as illustrated in  Figure~\ref{fig:layer_gam_mu}(b),
where we plot the magnetic moment (orange band)
\begin{equation}
\label{eq:mu}
    \mu=m_e c^2\frac{\gamma_\perp^2-1}{2 \Blg},
\end{equation} 
as a function of time for all the particles highlighted in Figure~\ref{fig:layerevol}. In the equation above, $\mathcal{B}$ is computed along the particle trajectory (see also Figure~\ref{fig:layer_r_b}(b)).

Soon after the particles enter the plasmoid (dashed gray line), their magnetic moment is, to a good approximation, conserved; the median value for the particle population (solid red line) is almost constant, and the variance (indicated by the width of the colored band) is much smaller than at earlier times. From the invariance of the magnetic moment ($\mu = \mathrm{const}$) and the magnetic field increase with time (see Equation \eqref{eq:BL_vs_t}) it follows that
\begin{equation}
    \gamma_{\perp}\propto \Blg^{1/2}\propto t^{1/2},
\end{equation}
where we used $\kappa=3/4$ to obtain the scaling with time. In Figure~\ref{fig:layer_gam_mu}(b) we also plot the second adiabatic invariant of the particles (green band)
\begin{equation}
\label{eq:Jpar}
    J_{\parallel}=\oint p_{\parallel}dl_{\parallel} \propto \gamma\beta_{\parallel}\Rlg\approx
    \gamma_{\parallel}\Rlg,
\end{equation}
assuming that particles have $\beta_\parallel\approx 1$, and find that $J_{\parallel}\approx \mathrm{const}$ for $ct/r_L \gtrsim 300$. This conservation (not discussed in PS18) yields
\begin{equation}
    \gamma_{\parallel}\propto 1/\Rlg(t)\propto t^{3/4}.
\end{equation}
where we used Equation~\eqref{eq:rL_vs_t} with $\xi=1$ and $\kappa=3/4$. Thus, combining the conservation of the first two adiabatic invariants of the particles with the growth of the (Lagrangian) magnetic field strength, we can explain the scalings of $\langle \gamma_{\parallel}\rangle, \langle\gamma_{\perp}\rangle$ from our simulations (Figure~\ref{fig:layer_gam_mu}(a)). \retwo{For the the nonrelativistic version of these relations for $P_\perp(t)$ and $P_\parallel(t)$, we refer the reader to  \cite{2017PhPl...24f2906M}.} 

We discuss next the conditions for conservation of $\mu$ and $J_{\parallel}$ and check whether they are indeed satisfied in our simulations. The first adiabatic invariant is conserved if the particle gyration timescale ($\omega_g^{-1}$) is much shorter than the characteristic timescale for the change of the magnetic field ($\Blg/\dot{\Blg}$). The particle gyration timescale in the plasma shell (see blue ring in Figure~\ref{fig:layerevol}) can be written as
\begin{equation}
    \omega_g^{-1} = \frac{B_{\rm up}}{\Blg}\cdot
        \frac{\langle\gamma\rangle}{\sups}\cdot
        \frac{r_L}{c}
    \sim \mathcal{O}(1) \frac{r_L}{c},
\end{equation}
where we assumed that $\langle\gamma\rangle\sim\sups$ and $\Blg\sim B_{\rm up}$ (see Figures \ref{fig:layer_gam_mu} and \ref{fig:layer_r_b}(b), respectively).
The timescale for the change of the magnetic field can be written as
\begin{equation}
    \frac{\Blg}{\dot{\mathcal{B}}}\sim \mathcal{O}\left(10^2\right) 
        \frac{r_L}{c} \gg \omega_g^{-1},
\end{equation}
as suggested by the results shown in Figure~\ref{fig:layer_r_b}(b). Thus, the conservation of the first adiabatic invariant is satisfied inside the plasmoid shell. 

The second adiabatic invariant, also known as the {\it mirror invariant}, is conserved if the time to cross the system in the direction parallel to the magnetic field  is much shorter than the characteristic timescale for the change of that dimension ($\Rlg/\dot{\Rlg}$). The former timescale can be approximated by $2\pi \Rlg/c$, assuming that particles move along magnetic field lines with $\beta_\parallel \approx 1$. From Figure~\ref{fig:layer_r_b}(a), we find order-of-magnitude estimates for both timescales, which read 
\begin{equation}
    \frac{\Rlg}{\dot{\Rlg}}\sim \mathcal{O}\left(10^2\right) \frac{r_L}{c},
    ~\text{while}~
    \frac{\Rlg}{c} \sim \mathcal{O}(10) \frac{r_L}{c}.
\end{equation}
Thus, the second adiabatic invariant is also conserved within a typical isolated plasmoid.

Let us also check that the particles can indeed be confined within the plasmoid during the energization process. On average, the Larmor radius of particles descending in the plasmoid shell can be written as $\tilde{r}_L \sim (\langle\gamma\rangle/\sups)(B_{\rm up}/\Blg)r_L \sim \mathcal{O}(1)r_L$, which is much smaller than the plasmoid size, $\rups \sim \mathcal{O}(10\text{-}10^2) r_L$ (see  Figure~\ref{fig:layer_gam_mu}(a)). This suggests that most of the energetic particles are trapped within the plasmoid. In fact, as particles sink toward the plasmoid core, they get increasingly more magnetized;  their Larmor radii decrease, since the magnetic field strength grows faster than the particle energy, namely, $\langle\gamma\rangle\propto t^{1/2}$, whereas $\Blg\propto t$, which leads to $\tilde{r}_L\propto t^{-1/2}$.


\subsection{Comparison to analytical model}
\begin{figure*}[htb]
    \centering
    \includegraphics[width=2\columnwidth]{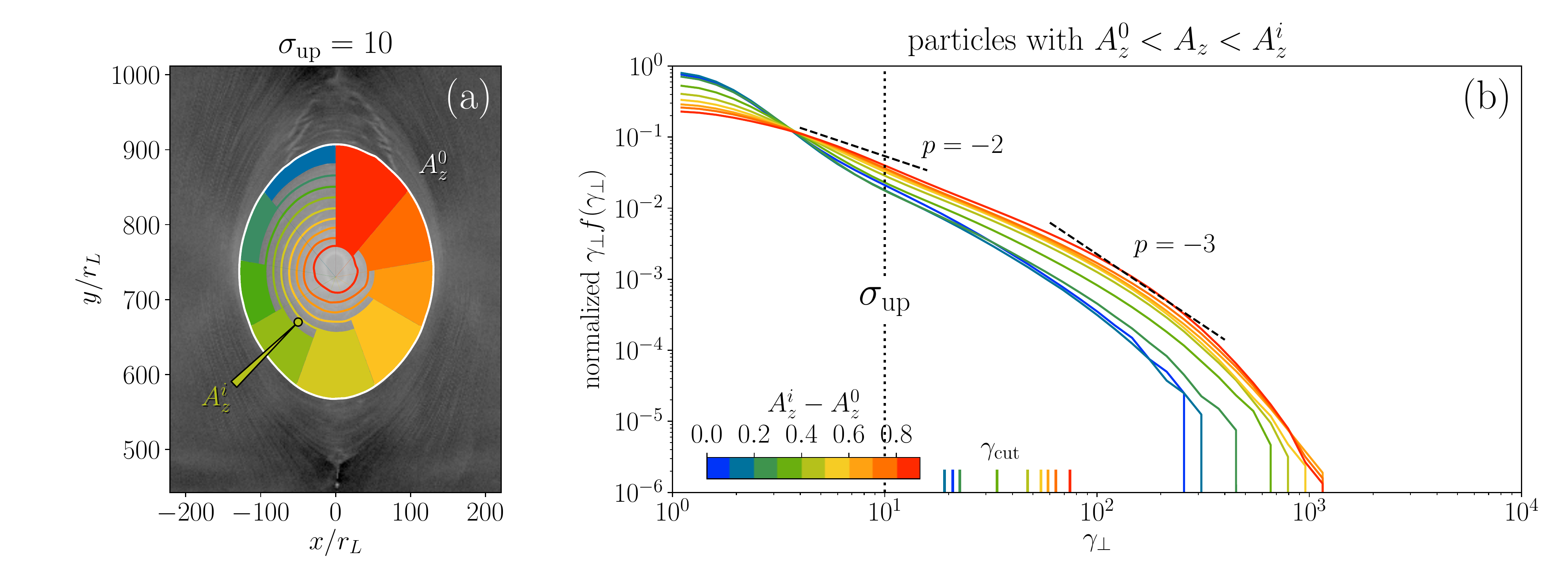}
    \includegraphics[width=2\columnwidth]{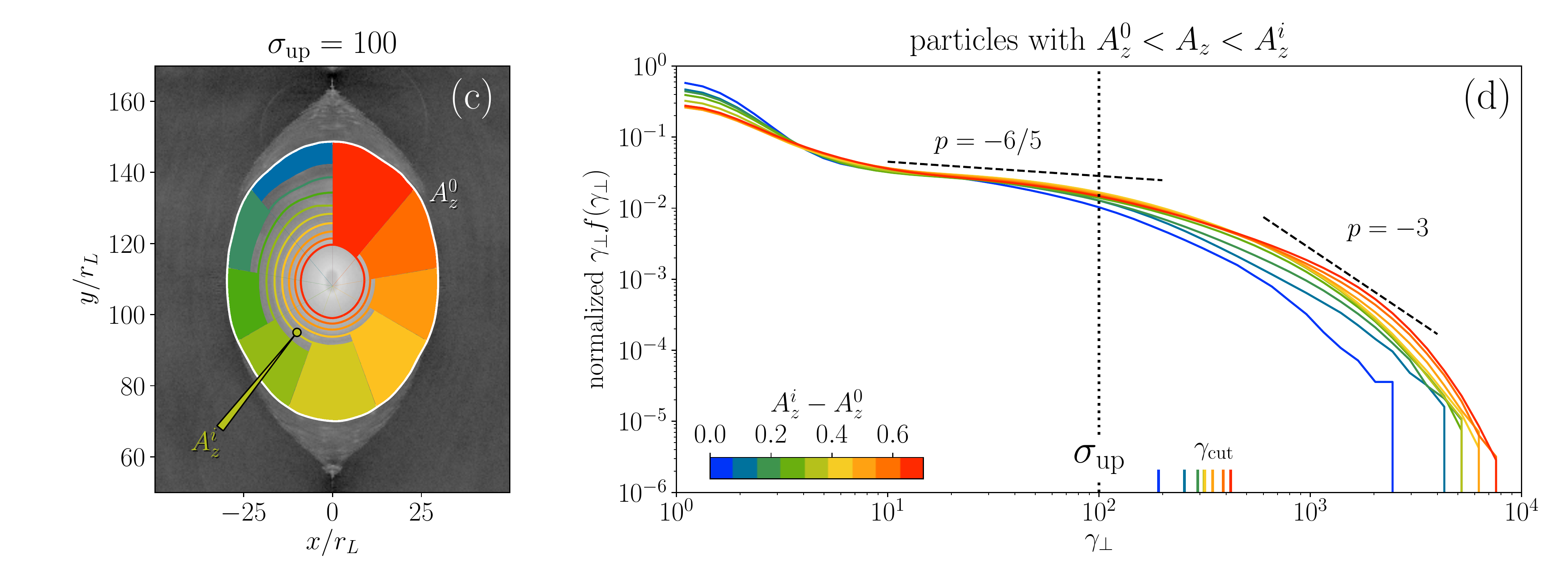}
    \caption{(a, c) Contours of the magnetic vector potential, $A_z$, are shown with colored lines. Rings with $A_z^0 < A_z < A_z^i$ are marked using colored sectors to guide the eye. Contours are overlaid on the plasma density shown in gray from our $\sups=10$ (top) and $100$ (bottom) simulations. The white line indicates the plasmoid boundary $A_z^0$. (b, d) Distribution functions (in perpendicular direction to the magnetic field, $f(\gamma_\perp)$) of particles belonging to different rings; blue curves correspond to the outermost region, while yellow/red curves are for the entire plasmoid (except for the innermost core). Colored vertical dashes represent the corresponding cutoff energies, computed as described in Section~\ref{sec:conserv_ad_inv}. Distribution functions are computed for the whole ring defined by $A_z^0 < A_z < A_z^i$.}
    \label{fig:layers_df}
\end{figure*}
In this section, we compare the predictions of our analytical model about the particle energy spectrum presented in Section~\ref{sec:analytics} with the results of two simulations of reconnection with $\sups=10$ and $\sups = 100$.


First, we select an isolated plasmoid and partition it into several concentric disks defined by equally spaced contours of the vector potential, $A_z$. This is shown in panels (a) and (c) of Figure~\ref{fig:layers_df}, where the $i$-th colored disk is defined by $A_z^0 < A_z < A_z^i$, and the white solid line represents the plasmoid boundary $A_z^0$. Disks are chosen in such a way that the $i$th disk contains all the $(i-1)$th, $(i-2)$th , ..., $0$th disks forming an onion structure, with the largest (red) disk containing all the other ones. We then pick particles from each of these disk regions and compute their distribution functions in the direction perpendicular to the local magnetic field, i.e., $f(\gamma_\perp)$. These are displayed in panels (b) and (d) of the same figure. For comparison purposes, all distribution functions are normalized to the total particle number of each region.

The distribution function of each disk is composed of multiple particle ``generations,'' namely, particles that were injected into the compressing plasmoid at different times. In general, particles from the outer regions (blue and green colored regions) have spent less time within the plasmoid than particles residing in the inner regions (orange and red colored regions). The difference in the particle residence time within the compressing plasmoid is reflected in the distribution functions extracted from different regions. This is illustrated in panels (b) and (d) of Figure~\ref{fig:layers_df}, where we see that the energy spectra of particles from regions closer to the plasmoid center (orange and red curves) are systematically shifted to higher Lorentz factors compared to the spectra of particles from the outer regions (blue curves). Similarly, the cutoff Lorentz factor (also marked with vertical dashes in both panels) of the distribution function from the outer regions is lower and closer to $\sups$, as expected for the injection particle spectrum (see also Figure~\ref{fig:injspec}). Moreover, the spectral shape is roughly the same among different disk regions, suggestive of an energization process that acts upon particles of all energies similarly, like adiabatic compression. 

The distribution functions from both simulations can be phenomenologically described as smooth broken power laws  with a high-energy exponential cutoff. The low-energy part of the spectrum has the slope of the injected energy spectrum, which depends on $\sups$. On the contrary, the high-energy part of the spectrum can be roughly described as a power law with $p\approx -3$ for both values of upstream magnetization. The steep power-law segment can be easily mistaken for an exponential cutoff because of the often limited energy range this spans. Without prior expectation for the existence of this steep power law, it is no surprise that previous studies did not report this. It is also worth emphasizing that for simulations with $\sups \lesssim 10$, where the X-point acceleration predicts steep particle spectra ($p<-2$), the robust identification of the $p=-3$ power-law tail is very difficult.

\subsection{Secondary energization timescale}



The timescale for magnetic field compression in Lagrangian terms is $\mathcal{B} / \dot{\mathcal{B}}$. Since $\mathcal{B} = B(\mathcal{R}(t), t)$, from Equation~\eqref{eq:radialstruct} we see that this timescale is also directly related to the timescale of plasmoid growth
\begin{equation}
    \label{eq:acctimescale}
    \frac{\mathcal{B}}{\dot{\mathcal{B}}} \sim \frac{\rups}{\dot{r}_{\rm up}} \sim \frac{\rups}{v_{\rm in}},
\end{equation}
where $v_{\rm in}$ is the reconnection rate (this result is similar to~\citealt{2016MNRAS.462...48S}). This timescale is also a proxy for the secondary acceleration timescale $t_{\rm sec} = \gamma / \dot{\gamma}$, as the secondary acceleration is a direct consequence of the compression of plasmoid interior. Relation~\eqref{eq:acctimescale} implies that the timescale for plasmoid compression (and thus secondary particle acceleration) becomes longer as the plasmoid grows over time. For the particular case shown in Figure~\ref{fig:layerevol}, where we can take $r_{\rm up}\sim 50r_L$ and $v_{\rm in}\sim 0.1 c$, we find that $t_{\rm sec}\sim 500r_L / c$, which roughly corresponds to the acceleration timescale inferred by Figure~\ref{fig:layer_gam_mu}(a) (characteristic timescale for the growth of $\langle\gamma\rangle$ from $\sim 200$ to $\sim 400$).

%
%
%
%

\section{Discussion}
\label{sec:discussion}

The standard picture for particle acceleration in relativistic reconnection is that particles get energized at the X-points and in the current sheet before being ultimately trapped inside  plasmoids, where they do not undergo any further energization. However, our study shows that this standard picture is not correct. In fact,  the dynamics of plasmoid compression is what actually dictates the formation of the overall spectrum in the long term, since most of the plasma in our 2D simulations ends up being trapped inside these islands. The potential drop across the X-points is limited to a few $\sups$, thus setting an upper bound (or a cutoff in energy) to which particles can get accelerated in astrophysical current sheets. Plasmoid compression, on the other hand, while being a slower process, can potentially accelerate particles to much higher energies.

\subsection{Pressure anisotropy}
\label{sec:anisotropy}
The energization process inside compressing plasmoids relies on the conservation of the first and second adiabatic invariants for magnetized particles trapped therein. As such, this process affects differently the parallel and perpendicular components of the particle momenta with respect to the local magnetic field. \retwo{This can cause a slow buildup of pressure anisotropy $P_{\parallel}/P_{\perp} \propto t^{1/4}$.} However, as discussed in Section~\ref{sec:part-evol}, on the scales of our simulations, we find that the pressure is almost isotropic inside the plasmoid shell, namely, $P_{\perp}\approx 2P_{\parallel}$. 

\retwo{While on timescales of our simulations it is impossible to identify an anisotropy buildup this subtle, particles may scatter off the small substructures present inside the plasmoids (see Figure~\ref{fig:plasm_example}). This process can slowly isotropize their distribution on timescales comparable to the plasmoid-crossing timescale $\mathcal{R}/c$. This timescale is much longer than the gyration period, and the conservation of the first adiabatic invariant will not be strongly affected by it. However, the conservation of the mirror invariant may be violated, in which case the evolution of $\gamma_\parallel$ will be dictated by the isotropy condition: $\langle\gamma_\parallel\rangle / \langle\gamma_\perp\rangle = 1/2$ (where averaging is done over a population of particles on a single field line). However, as it was pointed out in Section~\ref{sec:analytics}, this will affect the process only marginally, slightly enhancing the energization rate.}

Although not explicitly shown, we find deviations from isotropy (with $2P_{\parallel}<P_{\perp}$) in the plasmoid corona and plasmoid outskirts. While this evidence requires further analysis, it could cause the growth {\it of microinstabilities driven by pressure anisotropy} which tend to isotropize particle distribution functions \citep[e.g.,][]{2014PhRvL.112t5003K}. While these microinstabilities are typically studied for moderate plasma $\beta$ values ($\beta\gtrsim 1$), in our simulations we typically have $\beta\ll 1$ in plasmoid corona. This puts more stringent constraints on the development of these instabilities, requiring stronger pressure anisotropy.

\subsection{Possible limitations of the model} 
The efficiency of the secondary acceleration within plasmoids may vary in reconnecting systems with different physical conditions. In particular, the presence of a strong guide field can make plasmoids nearly incompressible, thus strongly interfering with this secondary acceleration process (J.~Davelaar, \& A.~Philippov, 2019 private communication). In 3D reconnecting systems the picture can also be different. Instead of 2D flux loops the plasmoids in 3D will look like elongated tubes that might further break into ellipsoids owing to the kink instability \citep[e.g.,][]{2011PhPl...18e2105L, 2014ApJ...783L..21S}. The motion and trapping of particles in these magnetic structures will be different than in their 2D counterparts studied here, as the highest-energy particles might be able to leave plasmoids \citep[e.g.,][]{2019ApJ...884..118L}. In turbulent magnetized plasmas, current sheets and plasmoids may not live long enough for the slow secondary energization process to develop, as plasmoids get stochastically formed and disrupted on timescales much shorter than the secondary energization timescale, $t_{\rm sec}$ \citep[e.g.,][]{2017PhRvL.118e5103Z, 2018PhRvL.121y5101C}.


Radiative cooling (due to synchrotron or inverse Compton scattering) may halt the secondary energization process in certain systems (see \citealt{2018JPlPh..84c7501N, 2019ApJ...877...53H, 2019MNRAS.482L..60W}). To quantify the effect of radiative cooling, we define a ``saturation'' Lorentz factor for which the secondary acceleration timescale (see Equation~\eqref{eq:acctimescale}) is comparable to the cooling timescale.

The cooling timescale for particles with Lorentz factor $\gamma$ can be estimated from the following relation:
\begin{equation}
    t_{\rm cool} \sim \frac{m_e c}{\sigma_T U \gamma},
\end{equation}
where for synchrotron cooling $U=B^2/8\pi$ (here $B$ is the average magnetic field strength of the plasmoid), while for inverse Compton cooling (in the Thomson regime) $U$ is the energy density of the background soft photon bath. Comparing this to the secondary acceleration timescale from Equation \eqref{eq:acctimescale} we find
\begin{equation}
    \label{eq:gammacrit}
    \gamma_{\rm sat} \sim \frac{m_e c v_{\rm in}}{w U \sigma_T},
\end{equation}
where $w$ is the characteristic plasmoid size (i.e., half-width in the perpendicular direction of the plasmoid motion). The value of $\gamma_{\rm sat}$ gives us a rough estimate of the Lorentz factor to which the secondary acceleration can energize electrons or positrons in a reconnecting current sheet, given the limitation from the radiative cooling. 

The fragmentation  of the reconnection layer results in the formation of plasmoids of different sizes, ranging from a few plasma skin depths to a sizable fraction of the layer's length $L$ \citep[e.g.,][]{2007PhPl...14j0703L, 2010PhRvL.105w5002U, 2016MNRAS.462...48S}, as also illustrated in Figure~\ref{fig:evolution}. The impact of secondary energization on the plasmoid chain will differ, as the energization timescale depends on the plasmoid size (i.e., smaller plasmoids contract faster; see Equation~\ref{eq:acctimescale}). In the following subsection, where we discuss the astrophysical implications of our results, we will consider for simplicity a typical large plasmoid that forms in the layer with size $w\sim 0.1 \, L$. Such large plasmoids contain most of the radiating particles of the layer and can have a significant contribution to the radiation emerging from the layer \citep[see, e.g.,][]{2016MNRAS.462.3325P,2018MNRAS.475.3797P}.

When the radiative cooling is not limiting the secondary acceleration, the maximum energy is determined by how large can plasmoids grow given the size and geometry of the source, how consistently the system can provide fresh plasma and magnetic flux, and how long the energetic particles can be constrained inside the plasmoid.


\subsection{Astrophysical implications} 

There are several astrophysical systems where the long-term acceleration scenario could play a role in shaping the energy distribution of radiating particles and producing a broken power law. In this model the acceleration time is assumed to be much smaller than the system lifetime, which is of order of few $L/c$. The only limiting factor in this case is the radiative cooling, which will effectively set the maximum energy to which particles can be accelerated via secondary energization.  In the following paragraphs, we make a qualitative discussion about the secondary acceleration and the possible impact of radiative cooling in jetted active galaxies, pulsars, and accretion disk coronae around black holes.


In {\it coronae of accretion disks around black holes} reconnection of magnetic flux tubes has been proposed to produce the nonthermal emission of the hard state of X-ray binaries. Relativistic particles in these systems are cooled on very short timescales (via either synchrotron or inverse Compton emission), much faster than the secondary energization. This conclusion is consistent with the results of \citet{2017ApJ...850..141B, 2020ApJ...899...52S}, and  \citet{2019MNRAS.482L..60W}, where emission models are based on bulk motions of cooled plasmoids instead of nonthermal acceleration.

Reconnection has also been shown to take place in the outer magnetospheres of {\it pulsars}, near the so-called light cylinder, producing nonthermal particle populations, which then emit pulsed synchrotron emission in X-rays and $\gamma$-rays \citep{1996A&A...311..172L, 2012MNRAS.424.2023P, 2014ApJ...780....3U, 2016MNRAS.457.2401C, PSAS18}. Typical sizes of the largest plasmoids in this scenario can be assumed to be equal to some fraction of the light cylinder radius, $w\sim 0.1 \, R_{\rm LC}$, where $R_{\rm LC}\sim c P / 2\pi$ and $P$ is the rotation period of the pulsar. The magnetic field decays as $r^{-3}$ from the neutron star surface ($B_*$) to the light cylinder. By equating the secondary energization rate with the synchrotron cooling rate, we find for regular pulsars
\begin{equation}
    \gamma_{\rm sat}\sim 3\times 10^2 
        \left(\frac{w}{0.1\, R_{\rm LC}}\right)^{-1}
        \left(\frac{P}{0.1~\text{s}}\right)^5,
\end{equation}
and for millisecond pulsars
\begin{equation}
    \gamma_{\rm sat}\sim 10^4 
        \left(\frac{w}{0.1\, R_{\rm LC}}\right)^{-1}
        \left(\frac{P}{5~\text{ms}}\right)^5,
\end{equation}
where we adopted $B_*\sim 10^{12}$ G for regular pulsars and $B_*\sim 10^8$ G for millisecond pulsars. Regular pulsars typically have magnetization parameters of $\sups\sim 10^3\text{-}10^5\gg \gamma_{\rm sat}$ close to the light cylinder \citep{1990ApJ...349..538C}. In other words, synchrotron cooling in this case is so strong that it limits the secondary acceleration to Lorentz factors well below $\sups$. In these systems, the formation of the nonthermal tail at $\gamma \gtrsim \sups$ is hampered by the radiative cooling. However, in millisecond pulsars, where the predicted $\gamma_{\rm sat}$ is an order of magnitude higher than for regular pulsars, the cooling might be slow enough for particles with energies $\sups \lesssim \gamma \lesssim \gamma_{\rm sat}$ for the secondary acceleration to matter.


In {\it blazars} -- active galaxies with relativistic jets closely aligned to the line of sight -- reconnection is believed to take place either in Poynting-flux-dominated jets \citep[e.g.,][]{2019MNRAS.484.1378G} or directly in the highly magnetized regions of accretion disks feeding the supermassive black holes, or close to the interface of jets and accretion flows \citep[e.g.,][]{2010A&A...518A...5D}. The occurrence of such reconnecting regions was also confirmed in general relativistic MHD simulations of black hole accretions disks \citep{2020MNRAS.495.1549N, 2020ApJ...900..100R}. Reconnection is also thought to take place in the collimated magnetically dominated outflows (jets) launched by the black hole and/or by the inner parts of the accretion disk \citep{2006A&A...450..887G, 2017MNRAS.469.4957B, 2018MNRAS.474.3535G}. Plasmoids produced during this process have also been invoked to explain high-energy flaring emission from blazar jets \citep[e.g.,][]{2013MNRAS.431..355G, 2016MNRAS.462.3325P, 2019MNRAS.482...65C}.

The high-energy radiation from blazar sources (i.e., from X-ray to $\gamma$-ray energies)  is usually modeled as inverse Compton emission by a nonthermal population of relativistic electrons and positrons in the jet scattering off low-energy photons (these can be synchrotron photons produced by the same particles or can originate from a radiation source external to the jet). The characteristic maximum energy to which particles are being accelerated sets a lower bound for the plasma magnetization parameter, which limits particle pre-acceleration in reconnection. Most blazar radiation models rely on a broken power-law distribution of injected particles to explain the observed broadband spectra \citep[e.g.,][]{2008MNRAS.385..283C, 2010MNRAS.401.1570T, 2013ApJ...768...54B}. For blazars typically the break occurs at $\gamma_{\rm b}\sim 10^2\text{-}10^3$, while the assumed nonthermal distribution usually spans up to $\gamma_{\rm max}\sim 10^5\text{-}10^6$. The power-law index typically varies from $p\sim [-1,-2]$ for energies below the break to $p\sim [-3,-4]$ for energies above the break. 

The secondary energization process described in this paper naturally produces a broken power-law distribution of particles. Moreover, the power-law index below the break, which is determined by the upstream plasma magnetization (see Figure~\ref{fig:injspec}), is similar to the values inferred by radiation modeling, for $\sups \gg 1$; for $\sups \sim 10$, $p\sim -2$, while for $\sups \gtrsim 10$, $p\sim [-1.5,-1]$ \citep[e.g.,][]{2014PhRvL.113o5005G, 2014ApJ...783L..21S,  2016ApJ...816L...8W}. Meanwhile, the plasmoid compression leads to a spectral break at $\gamma_{\rm b} \approx \mathcal{O}(1)\sups$ (which resembles the observed $\gamma_{\rm b}$ assuming $\sups\sim 10^2\text{-}10^3$) and to an asymptotic power-law index $p\approx -3$ above the break, in agreement with the radiation models. Moreover, the maximum energy reached by the particles due to the plasmoid compression is not limited by the magnetization; thus particles can, in principle, reach $\gamma_{\max} \gg \sups$.


For these systems, assuming that the leading cooling mechanism is synchrotron emission, we find a conservative estimate for the ``saturation'' energy defined by Equation~\eqref{eq:gammacrit}
\begin{equation}
    \gamma_{\rm sat}\gtrsim 
        10^5 \left(\frac{w}{10^{14}~\text{cm}}\right)^{-1}
                \left(\frac{B}{1~\text{G}}\right)^{-2},
\end{equation}
where  $w\sim 0.1\, L$ is the typical plasmoid size, $B\sim 1$ G is the magnetic field within the plasmoid, and the typical size of the system $L\sim 10^{15}$ cm \citep[see, e.g.,][]{2008MNRAS.385..283C}. This estimation is close to the typical maximum injection energy, $\gamma_{\rm max}$, assumed when modeling the radiation from these systems. Thus, the secondary energization in plasmoids is a plausible mechanism for producing the broken power-law distributions of radiating particles in blazars.




\section{Summary}
\label{sec:summary}
Fast magnetic reconnection is accompanied by the formation of a self-similar chain of plasmoids, which accumulate particles both from the adjacent current sheets and directly from the upstream region. Because of this constant accretion of particles and magnetic flux, plasmoids grow in size, while their interiors get compressed, as particles are advected inward closer to the plasmoid core by converging magnetic flux loops. The radial structure of these plasmoids is independent of the upstream conditions and is determined exclusively by the force balance between the magnetic stresses and the plasma pressure.

We find that the highest-energy particles in our 2D simulations typically undergo a two-stage acceleration during their lifetime. They first get energized in the current sheets and X-points; this process forms the initial power-law distribution function, which depends on the upstream magnetization, $\sups$. These particles are then advected into plasmoids. As particles are advected toward the plasmoid center by the converging field lines, they experience an almost linearly growing magnetic field with time, while their adiabatic invariants are roughly conserved. As a result, $\gamma\sim \gamma_\perp \propto t^{1/2}$ and $\gamma_{\parallel} \propto t^{3/4}$, with $\gamma_\perp \sim 2\gamma_{\parallel}$, i.e., the pressure is isotropic. 



The power-law slope of injected particles, which depends on $\sups$, is conserved and extends to Lorentz factors of a few $\sups$ \citep{2016ApJ...816L...8W}. Meanwhile, a second power law with slope $p\approx -3$ forms at $\gamma\gtrsim \mathcal{O}(1)\sups$, and is followed by a time-evolving high-energy cutoff, $\gamma_{\rm cut}\propto t^{1/2}$. The rate of this secondary energization is primarily dictated by the large-scale reconnection dynamics and is independent of the upstream parameters. For a plasmoid with transverse width $w$, the secondary energization timescale is $\gamma/\dot{\gamma} \sim w / v_{\rm in}$, where $v_{\rm in}\sim 0.1 c$ is the global reconnection rate. As the plasmoid grows over time, the secondary energization will become slower. Ultimately the particle energization will cease, and $\gamma_{\rm cut}$ will stop growing, once the secondary acceleration timescale becomes comparable to the radiative cooling timescale for a given astrophysical system. We find that in the outer magnetospheres of millisecond pulsars and reconnecting regions in blazar jets the cooling may be weak enough for this slow secondary process to accelerate particles beyond the standard $E\sim \sups m_e c^2$ limit, forming an additional power-law tail $E^{-3}$ at higher energies. 

Although the secondary energization process was studied for reconnection in pair plasmas, we argue that it can operate also in magnetically dominated electron-ion plasmas, since all species are accelerated to roughly the same energy, and the secondary acceleration due to plasmoid compression proceeds in the same way.

\begin{acknowledgements}
The authors would like to thank Alexander Philippov for numerous discussions and insightful comments. \retwo{The authors would also like to thank the anonymous reviewer for critical comments that helped to clarify certain points and improve the quality of this paper.} This research was supported in part by the National Science Foundation under grant No. NSF PHY-1748958, NASA ATP grant No. 80NSSC18K1099 and NSF grant AST-1814708. M.P. acknowledges support from the Lyman Jr.~Spitzer Postdoctoral Fellowship and the Fermi Guest Investigation grant No.~80NSSC18K1745. A.S. is supported by the Simons Foundation (grant 267233). L.S. acknowledges support from the Sloan Fellowship, the Cottrell Fellowship, NASA ATP NNX17AG21G, and NSF PHY-1903412. 
\end{acknowledgements}

\appendix
\counterwithin{figure}{section}

%
%
%
%

\section{Structure of the plasmoid shell}
\label{appendix:plasm}
This appendix focuses on the internal structure of primary isolated plasmoids. We estimate the power-law indices, defined by Equation~\eqref{eq:radialstruct}, of the radial profiles of the magnetic field and plasma density inside the plasmoid shell, $\rins < r < r_0(t)$. We also derive how the distance of particles from the plasmoid center decreases with time, as particles slowly descend toward it.

First, let us assume that at any given radius from the center of the plasmoid there is a balance between the magnetic forces and plasma pressure
\begin{equation}
    \label{eq:forcebalance}
    \frac{1}{c}\bm{j}\times\bm{B} = \nabla P,
\end{equation}
where the current density $\bm{j}$ can be expressed as $4\pi\bm{j}/c=\nabla\times\bm{B}$. Motivated by the simulation results, we assume that, within the plasmoid shell, $\bm{B}$ is purely toroidal and the only variation occurs in the radial direction, i.e., $\bm{B}=B(r)\bm{\hat{\phi}}$. Then, Equation~\eqref{eq:forcebalance} can be rewritten as
\begin{equation}
    \label{eq:b1}
    -\frac{B^2}{r} = \frac{d}{dr}\left(4\pi P + \frac{B^2}{2}\right) \cdot
\end{equation}

We also assume a polytropic EOS for the plasma inside the plasmoid shell, with isotropic pressure
\begin{equation}
    \label{eq:state}
    P = K\rho^\Gamma,
\end{equation}
where $K$ is some dimensional constant and $\Gamma$ is the adiabatic index. Substitution of Equations~\eqref{eq:radialstruct} and  \eqref{eq:state} into Equation~\eqref{eq:b1} yields
\begin{equation}
\label{eq:b2}
    \frac{(1-\zeta)\sigma_0}{\xi(\Gamma-1)} = \left(\frac{r}{r_0(t)}\right)^{2\zeta-\xi\Gamma},
\end{equation}
where $\sigma_0$ is the plasma magnetization at $r=r_0$. This can be expressed as
\begin{equation}
    \label{eq:sigma0}
    \sigma_0 \approx \frac{B_0^2(\Gamma-1)}{4\pi \Gamma K\rho_0^\Gamma},
\end{equation}
where we used the definition for the plasma magnetization $\sigma_0 = B_0^2/4\pi h_0$, and expression of the enthalpy density $h_0$ (at $r=r_0$) for a relativistically hot plasma ($kT_0 \gg m_e c^2$)
\begin{equation}
    h_0 = \rho_0 c^2\left(1 + \frac{\Gamma}{\Gamma-1} \frac{kT_0}{m_e c^2}\right) \approx \frac{\Gamma}{\Gamma-1} P_0,
\end{equation} 

For Equation~\eqref{eq:b2} to be satisfied at all times and for all $\rins <r\le r_0$, the following relations must hold:
\begin{equation}
    2\zeta=\xi\Gamma,~\text{and}~(1-\zeta)\sigma_0=\xi(\Gamma-1).
\end{equation}
Solving the above equations for the unknown power-law indices $\zeta$ and $\xi$, we find
\begin{equation}\label{eq:powerlaws_app}
    \zeta=\frac{\Gamma\sigma_0/2}{\Gamma + \Gamma\sigma_0/2 - 1},~\text{and}~\xi = \frac{\sigma_0}{\Gamma + \Gamma\sigma_0/2 - 1} \cdot
\end{equation}

Particles inside the plasmoid shell are frozen into the slowly contracting magnetic field loops, which bring the particles closer to the plasmoid center. As a result, the mass enclosed within a fixed magnetic loop in the plasmoid shell is approximately constant in time. This condition can be expressed as
\begin{equation}
    \int_{\rins}^{\Rlg}r \rho(r, t)dr \approx \mathrm{const},
\end{equation}
where $\Rlg$ is the decaying radius of a fixed magnetic loop or plasma ring (see Figure~\ref{fig:layerevol}). 
This condition, together with Equations~\eqref{eq:radialstruct} and \eqref{eq:inflationrate}, yields
\begin{equation}
    \label{eq:lagrangian_radius}
    \Rlg\propto r_0(t)^{-\xi/(2-\xi)} \propto t^{-\kappa\xi/(2-\xi)}
\end{equation}
where we assumed $\rins \ll \Rlg$.

\begin{figure*}[htb]
    \centering
    \includegraphics[width=2\columnwidth]{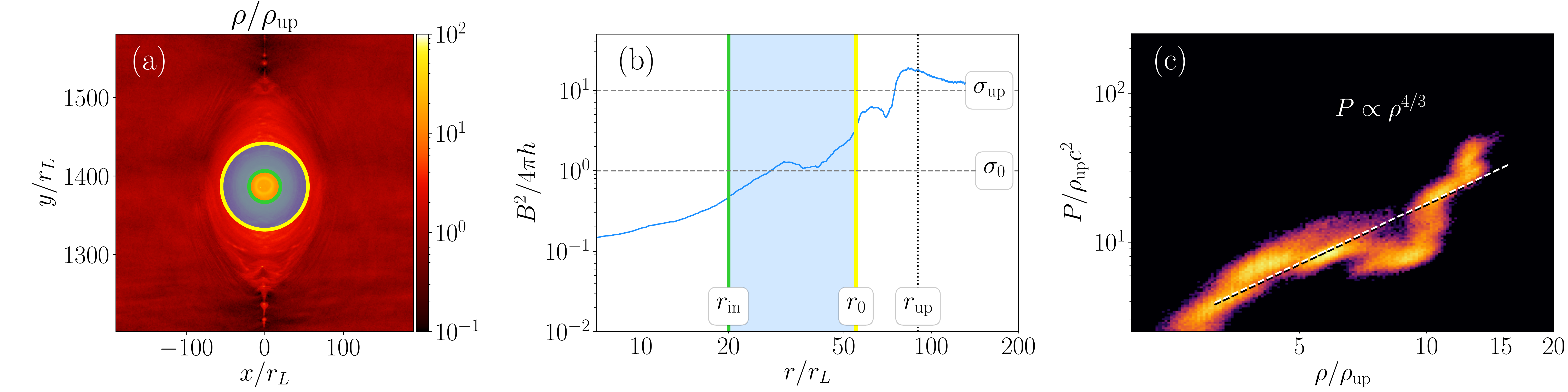}
    \includegraphics[width=2\columnwidth]{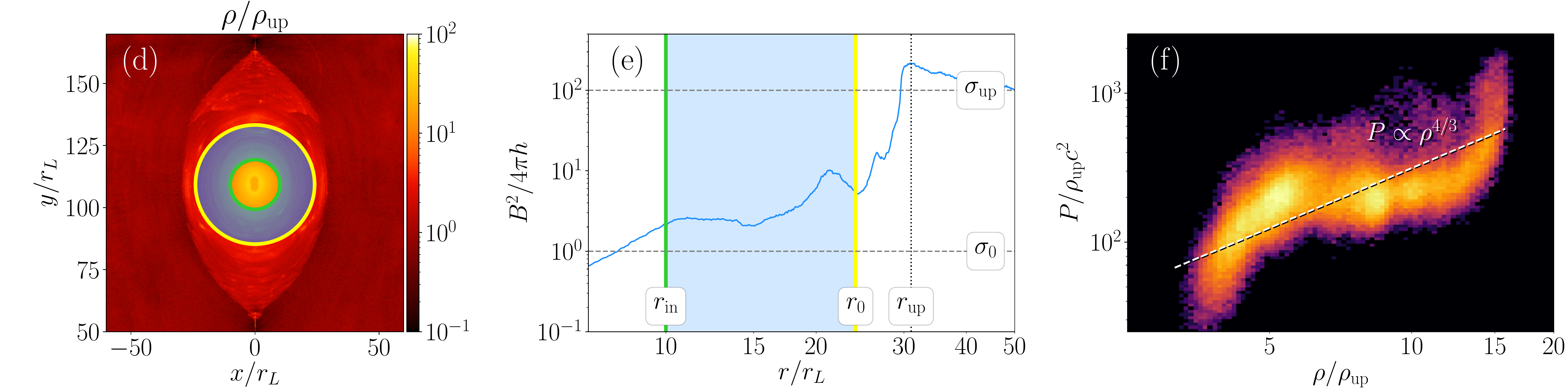}
    \caption{Top panels correspond to the $\sups=10$ simulation, while bottom panels are for the $\sups=100$ case. (a, d) Close-up view of an isolated primary plasmoid, with color indicating $\rho/\rho_{\rm up}$ (see color bar). The plasmoid shell, where the force balance condition \eqref{eq:forcebalance} is satisfied, is shown as a blue shaded ring; the green and yellow circles indicated the core radius and the boundary of the shell. (b, e) Plasma magnetization as a function of radius from the plasmoid center (in units of $r_L$). The blue shaded region corresponds to the plasmoid shell, shown in blue in panels (a) and (d). The horizontal dashed lines correspond to the upstream magnetization, $\sups$, and the magnetization of the plasmoid shell, $\sigma_0$ (see Equation~\eqref{eq:sigma0}). The three radii marked on the plot are defined in Section~\ref{sec:plasmoids}. (c, f) Typical two-dimensional histogram of the plasma pressure and plasma density for the shaded region in panels (a) and (d). The polytropic EOS for a relativistic gas is also shown (dashed line).}
    \label{fig:plasm_eos}
\end{figure*}

As an example, Figure~\ref{fig:plasm_eos} shows results from our simulations (for a description, see Section~\ref{sec:setup}) for $\sups=10$ (top row) and $100$ (bottom row). Panels (a) and (d) show the region of the plasmoid where the force balance is satisfied, panels (b) and (e) show magnetization as a function of radius from the plasmoid center (blue shaded region corresponds to the same region in (a) and (d)), and panels (c) and (f) show the EOS for the same region (top and bottom panels correspond to different upstream magnetizations, $\sigma_{\rm up}=10$ and $\sigma_{\rm up} = 100$). As we see from panels (b) and (e), the effective magnetization drops from the upstream value to a roughly constant value $\sigma_0\approx 1$ in the plasmoid shell. From panels (c) and (f) we can see that the EOS indeed looks like a polytrope with a characteristic adiabatic index of $\Gamma=4/3$. 

Thus, for $\Gamma=4/3$ and $\sigma_0\approx 1$ from Equation \eqref{eq:powerlaws_app} we find that $\zeta\approx 2/3$ and $\xi \approx 1$. From Equation \eqref{eq:lagrangian_radius} we also find that $\Rlg\propto t^{-\kappa}$ when $\xi\approx 1$.

%
%
%
%

\section{Finding the boundaries of plasmoids}
\label{appendix:plasmbound}
In this section, we describe the algorithm we used for identifying the plasmoid boundaries. This relies on the mixing criterion~\citep{2014PhPl...21e2307D, 2017ApJ...850...29R} and on the vector potential.

We distinguish particles originating from one side of the current sheet, $+x$, from the ones from the other side, $-x$. Henceforth, we refer to their densities as $\rho^+$ and $\rho^-$. We then compute the so-called {\it mixing factor}, $\lambda_f$, in each cell of our simulation domain

\begin{equation}
    \lambda_f = 1 - \left(1 - 2\frac{\rho^+}{\rho^+ + \rho^-}\right)^2.
    \label{eq:mixingf}
\end{equation}

\begin{figure*}[htb]
    \centering
    \includegraphics[width=1.5\columnwidth]{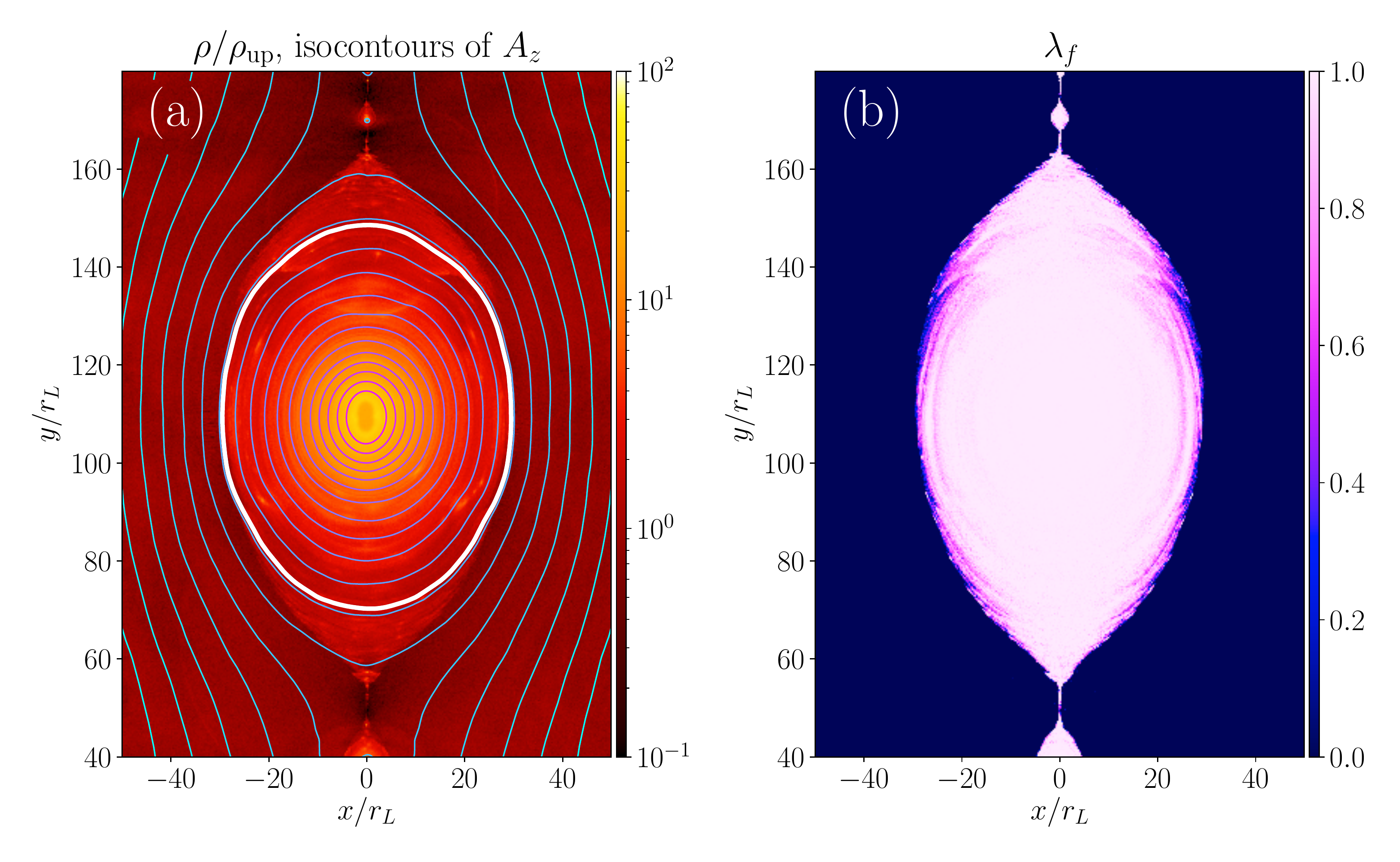}
    \caption{Zoom in of a small region of the reconnection layer (from the $\sups=100$ simulation) containing an isolated large plasmoid. Panel (a) Plasma density in logarithmic scale (see color bar on the side)  with overlaid contours of the vector potential (solid cyan lines). The thick white line corresponds to $A_z=A_z^0$ and marks the plasmoid boundary. (b) Mixing factor, $\lambda_f$, defined by Equation~\ref{eq:mixingf}. The transition from $\lambda_f=1$ (fully mixed) to 0 (not mixed) happens over only a few skin depths.}    \label{fig:plasmbound}
\end{figure*}

The mixing factor is defined in a way that $\lambda_f = 1$ inside the plasmoids and the current sheet, where particles from two separated regions are perfectly ``mixed,'' and $\lambda_f = 0$ everywhere else. At the plasmoid edges the mixing factor takes intermediate values, $0 < \lambda_f < 1$ (see Figure~\ref{fig:plasmbound}(b)). We compute the isocontours of the vector potential $A_z$ (the simulation is done in the $x$-$y$ plane). To identify the boundary of a particular plasmoid, we select regions characterized by intermediate values of the mixing factor (i.e.,  $0.1 < \lambda_f < 0.9$) and find the average value of the vector potential values in these regions, $A_z^0$. We then define the isocontour of $A_z=A_z^0$ as the boundary for that particular plasmoid (see Figure~\ref{fig:plasmbound}(a), thick white line).

Our results are robust to the choice of the exact mixing factor values, as $\lambda_f$ has a very steep spatial profile at the plasmoid edges; it changes quickly from $0$ to $1$ going from the upstream to the plasmoid within a few skin depths, meaning that the mixing of particles happens very abruptly. Even if one argues that our method does not yield the exact plasmoid boundary, this does not affect our results, because our analysis focuses on long-term processes taking place well within the plasmoid boundary.





\bibliography{main}{}



\end{document}
